\documentclass[a4paper,11pt]{article}
\usepackage{graphicx,amssymb,bm,latexsym,color,epsf}
\makeatletter
\@addtoreset{equation}{section}

\makeatother
\newcommand{\be}{\begin{equation}}
\newcommand{\ee}{\end{equation}}
\newcommand{\bea}{\begin{eqnarray}}
\newcommand{\eea}{\end{eqnarray}}
\newcommand{\vs}[1]{\vspace{#1 mm}}
\newcommand{\hs}[1]{\hspace{#1 mm}}
\renewcommand{\a}{\alpha}
\renewcommand{\b}{\beta}
\renewcommand{\c}{\gamma}
\newcommand{\G}{\Gamma}
\renewcommand{\d}{\delta}

\newcommand{\s}{\sigma}
\renewcommand{\t}{\theta}

\newcommand{\la}{\lambda}
\newcommand{\pa}{\partial}

\newcommand{\nn}{\nonumber\\}
\newcommand{\p}[1]{(\ref{#1})}

\newcommand{\br}{\bar R}
\newcommand{\bR}{\bar R}
\newcommand{\bg}{\bar g}
\newcommand{\eom}{\bar E}

\newcommand{\bnabla}{\bar\nabla}

\newcommand{\Tr}{{\rm Tr}}
\newcommand{\Det}{{\rm Det}}
\def\lich{{\Delta_L}}
\def\zgf{Z_{GF}}

\newcommand\bgamma{\bar\gamma}

\begin{document}

\begin{titlepage}

\renewcommand{\thefootnote}{\fnsymbol{footnote}}
\begin{flushright}
KU-TP 068 \\
\today
\end{flushright}

\vs{10}
\begin{center}
{\Large\bf Gauges and functional measures in quantum gravity II:\\
Higher derivative gravity}
\vs{15}

{\large
N. Ohta,$^{a,}$\footnote{e-mail address: ohtan@phys.kindai.ac.jp}
R. Percacci$^{b,c,}$\footnote{e-mail address: percacci@sissa.it}
and A. D. Pereira$^{d,}$\footnote{e-mail address: duarte763@gmail.com}
} \\
\vs{10}

$^a${\em Department of Physics, Kindai University,
Higashi-Osaka, Osaka 577-8502, Japan}

$^b${\em International School for Advanced Studies, via Bonomea 265, 34136 Trieste, Italy}

$^c${\em INFN, Sezione di Trieste, Italy}

$^d${\em UERJ $-$ Universidade do Estado do Rio de Janeiro,
Departamento de F\'isica Te\'orica, Rua S\~ao Francisco Xavier 524,
20550-013, Maracan\~a, Rio de Janeiro, Brasil}

\vs{15}
{\bf Abstract}
\end{center}

We compute the one-loop divergences in a higher-derivative theory of gravity including Ricci tensor squared and Ricci
scalar squared terms, in addition to the Hilbert and cosmological terms,
on an (generally off-shell) Einstein background.
We work with a two-parameter family of
parametrizations of the graviton field, 
and a two-parameter family of gauges.
We find that there are some choices of gauge or parametrization
that reduce the dependence on the remaining parameters.
The results are invariant under a recently discovered
``duality'' that involves the replacement of the densitized metric 
by a densitized inverse metric as the fundamental quantum variable.

\end{titlepage}
\newpage
\setcounter{page}{2}
\renewcommand{\thefootnote}{\arabic{footnote}}
\setcounter{footnote}{0}

\section{Introduction}

In a previous paper~\cite{I}, hereafter referred to as I, we have examined the properties of quantum General Relativity 
(GR - the theory containing
only terms up to second derivatives in the action) 
in a general four-parameter family of gauges and parametrizations.
In this paper, we would like to extend the analysis to 
Higher-Derivative Gravity (HDG).

By HDG, we will always mean the theory
of gravity based on the metric as fundamental variable and
an action that contains up to four derivatives.\footnote{There has been recently some progress
on theories that contain more than four, and possibly infinitely many
derivatives~\cite{Modesto:2011kw,Biswas:2011ar,Modesto:2014lga}.
These also deserve the name HDG, but we shall not consider them here.}
This theory is important because in four dimensions it
is power-counting renormalizable \cite{Stelle1}
and asymptotically free \cite{julve,ft1,avrabar}.
In spite of these appealing properties, it has never been
accepted as a viable fundamental theory for gravity,
because of the presence of ghosts in perturbation theory.
Over time, there have been many proposals trying to
circumvent this problem.
Among these let us mention here the following possibilities:
\begin{itemize}
\item
The mass of the ghost is not a fixed parameter but is rather subject to strong (quadratic)
running above the Planck threshold.
Then, the equation for the pole mass $m_{phys}^2=m^2(k=m_{phys})$
(where $m(k)$ is the running mass) may not have a solution \cite{julve,salam,floper4}.
\item
The ghost may be an artifact of expanding around the wrong vacuum.
The true vacuum of quadratic gravity (in the presence also of a Hilbert term) is not flat space
but rather a kind of wave with wavelength
of the order of the Planck length \cite{bonannoreuter1}.
\item The quadratic term is one of an infinite series and
the sum of the series is a function that has no massive ghost pole.
The ghost pole is an artifact of Taylor expanding this function to second order
(see the aforementioned papers on non-local gravity and also \cite{Shapiro:2015uxa}).
\end{itemize}
For further proposals see also \cite{Tomboulis,Mannheim:2006rd,Mukohyama:2013ew}.
So far, none of these arguments has convinced the community at large, 
so the issue of the ghosts remains open for the time being.
However, the hope that a way out may exist has generated
new interest in these theories in recent times,
also among particle physicists \cite{Salvio:2014soa,Einhorn:2014gfa,AKKLR}.
It seems therefore appropriate to keep investigating the 
quantum properties of these theories.

In this paper we will extend previous results in several directions.
Due to the complicated structure of the theory,
calculations of one-loop divergences in HDG have usually 
been performed with a special four-derivative gauge-fixing term 
such that the four-derivative part of the Hessian is 
proportional to the square of the Laplacian.
In this paper we will calculate the off-shell gauge-dependence 
of the one-loop divergences by using the more conventional
second-derivative gauge-fixing term that
is commonly used in quantum GR, depending on two 
parameters $a$ and $b$, or a four-derivative variant
of the same gauge fixing, containing an extra power of a Laplacian.

As another generalization, we assume that the 
quantum field is not
the metric but the densitized metric
\be
\gamma_{\mu\nu}=g_{\mu\nu}\left(\sqrt{\det g_{\mu\nu}}\right)^w\ ,
\label{denslin}
\ee
or densitized inverse metric
\be
\gamma^{\mu\nu}=g^{\mu\nu}\left(\sqrt{\det g^{\mu\nu}}\right)^{-w}
=g^{\mu\nu}\left(\sqrt{\det g_{\mu\nu}}\right)^w\ ,
\label{densinv}
\ee
with weight $w$.
Furthermore we allow the quantum theory to depend on another
parameter $\omega$ that interpolates continuously between
the linear background-field expansion (for $\omega=0$)
\be
\label{linexp}
\gamma_{\mu\nu}=\bar\gamma_{\mu\nu}+\hat h_{\mu\nu}\ ,
\ee
the exponential background-field expansions (for $\omega=1/2$)
\be
\label{expexp}
\gamma_{\mu\nu}=\bar\gamma_{\mu\rho}(e^{\hat h})^\rho{}_\nu
\qquad \mathrm{or}\qquad
\gamma^{\mu\nu}=(e^{-\hat h})^\mu{}_\rho\bgamma^{\rho\nu}\ ,
\ee
and the linear expansion around the inverse densitized metric
(for $\omega=1$)
\be
\label{invexp}
\gamma^{\mu\nu}=\bar\gamma^{\mu\nu}-\hat h^{\mu\nu}\ .
\ee
The density $\hat h_{\mu\nu}$ is then reexpressed
as
\be
\hat h_{\mu\nu}=\left(\sqrt{\det\bg}\right)^w h_{\mu\nu}
\quad\mathrm{or}\quad
\hat h^{\mu\nu}=\left(\sqrt{\det\bg}\right)^w h^{\mu\nu}
\ee
respectively,
where $\bg_{\mu\nu}$ is a background metric, related to
the background density $\bgamma_{\mu\nu}$ as in (\ref{denslin},\ref{densinv}).
See also equation (\ref{redefh}) below.
The field $h_{\mu\nu}$ is used as integration variable in the
functional integral.
The properties of quantum GR in this general four-parameter family of gauges and parametrizations
have been investigated in an accompanying paper~\cite{I}.

Here we extend the analysis of I to HDG.
We now have four independent couplings instead of two, so the 
expressions for the divergences are in general much more complicated
than in I and so is the interpretation of the results.
The expressions simplify somewhat, and are reported explicitly,
in two limits: the ``four derivative gravity (4DG) limit'' in which the
Einstein-Hilbert terms can be neglected relative to the
curvature squared terms, and the ``Einstein-Hilbert (EH) limit''
where the opposite holds.
In the EH limit, the action is the same as
the one considered in I, but the analysis that we perform here
differs in two respects: first, we choose a more general Einstein
background, instead of the maximally symmetric background of I.
This allows us to discriminate two divergent terms quadratic
in curvature, rather than a single one as in I.
Secondly, the cutoff in each spin sector is chosen to depend 
on the corresponding Lichnerowicz Laplacian, rather than the
Bochner Laplacian $-\nabla^2$ as in I.
Thus, comparison with I yields some information on the
cutoff-dependence of these non-universal results.
Still, we find that in this limit the qualitative picture is the same.
Similar calculations of divergent terms with different parametrizations
in four dimensions have been given in \cite{KP}.

In four dimensions and in the 4DG-limit, all divergences are universal,
i.e. independent of both gauge and parametrization.
In particular, the logarithmic divergences are related 
to the well-known universal beta functions of HDG \cite{avrabar}.
As in I, we find that all divergences are invariant under a 
``duality'' transformation that consists in the replacement 
\bea
\label{dudu}
\omega &\to& 1-\omega ,
\nonumber\\
w &\to& w+\frac{4}{d}\ .
\eea

Let us discuss a little more the role of the functional measure
in these considerations.
In I we have used the word ``measure'' synonymously with
``choice of quantum field in the functional integral''
but we have left the definition of the functional measure 
a bit implicit.
The reason for this is that the Wilsonian cutoff $k$ that we 
have used to calculate the divergences does not regulate
the divergences in the functional measure, if there are any,
and so the results would have been independent of this choice anyway.
The consistent interpretation of the results of I is that they
give the correct divergences of the functional integrals when the
measure is given by
\be
\Pi_x d h_{\mu\nu}(x)\ ,
\label{measure}
\ee
where $h_{\mu\nu}$ is the purely tensorial quantum field
defined as in (\ref{redefh}),
independently of $m$ and $\omega$.

Since the functional measure was always kept fixed,
the dependence of the results on the parameters $m$ and $\omega$
resulted entirely from the different forms of the Hessians,
which in turn was due to the different forms of the expansion 
of the action (see equations (\ref{gammaexp},\ref{deltag}) below).

If one decided to use the same expansion but a different
ultralocal measure, for example 
\be
\Pi_x d\hat h_{\mu\nu}(x)
=\Pi_x (\det\bg(x))^{w/2} d h_{\mu\nu}(x)
\label{altmeasure}
\ee
then the divergences would differ by terms of the form $\delta(0)$
times the volume.
Such terms would affect the power-law divergences coefficients.

We conclude this introduction by listing the contents
of the following sections.
Section 2 contains details of our calculations of one-loop
divergences.
In section 3 we give a formal proof (at the level of determinants)
that the on-shell effective action in $d=4$
is gauge-independent.
(In particular, the effective action of HDG in the 4DG-limit 
on an Einstein space is gauge-independent.)
The main results are presented in section 4.
In section 5, we discuss how the results can be obtained for $d=4$ conformal gravity.
In section 6, we point out that our results show that the duality found in our previous paper I
is valid in HDG as well.
Section 5 contains a discussion and conclusions.
In the appendix, we summarize the heat kernel coefficients for Lichnerowicz Laplacians
on an Einstein manifold.

\section{The one-loop effective action}

\subsection{The HDG actions and their equations of motion}

In this paper we will consider actions of the general form
\bea
S(g)=\int d^d x \sqrt{\mp g} 
\Big[\pm Z_N(R - 2 \Lambda) + \a R^2
+\b R_{\mu\nu}^2 
\Big] ,
\label{action}
\eea
where $Z_N=1/(16\pi G)$, $\Lambda$ is the cosmological constant,
$\alpha$, $\beta$ 
are the higher derivative couplings.
The upper sign refers to Minkowski signature, the lower one
to Euclidean signature.
This is not the most general HDG action,
because we omit a term $\c R_{\mu\nu\rho\sigma}^2$,
however in $d\leq 4$ this is not a very strong restriction,
because the ``Gauss-Bonnet'' or ``Euler'' combination
\bea
G=R_{\mu\nu\a\b}^2-4 R_{\mu\nu}^2+R^2 ,
\eea
is either zero (in $d=2,3$) or a total derivative (in $d=4$).
Using the identity for the Weyl tensor $C_{\mu\nu\rho\s}$:
\bea
C^2\equiv C_{\mu\nu\rho\sigma}C^{\mu\nu\rho\sigma} 
= R_{\mu\nu\a\b}^2-\frac{4}{d-2} R_{\mu\nu}^2+ \frac{2}{(d-1)(d-2)} R^2,
\eea
one has
\be
C^2=G+\frac{2(d-3)}{d-2}W\ ;\qquad
W=2R_{\mu\nu}^2-\frac{d}{2(d-1)}R^2\ .
\ee
Thus, modulo terms proportional to $G$,
we can replace $\a R^2+\b R_{\mu\nu}^2$ by
\be
\frac{1}{2\lambda}W+\frac{1}{\xi}R^2\ .
\ee
The integral of $W$,
like the integral of $C^2$, is Weyl-invariant in $d=4$.

The equations of motion are
\bea
\pm Z_N(G_{\mu\nu} +\Lambda g_{\mu\nu}) 
+ \a E_{\mu\nu}^{(1)} 
+ \b E_{\mu\nu}^{(2)}
=0,
\eea
where $G_{\mu\nu}$ is the Einstein tensor and
\bea
E_{\mu\nu}^{(1)} &=& 2RR_{\mu\nu}-2\nabla_\mu \nabla_\nu R
 + g_{\mu\nu}\Big(2\Box R -\frac12 R^2\Big), \nn
E_{\mu\nu}^{(2)} &=& 2R_{\mu\la}R_\nu^\la - 2 \nabla^\la \nabla_{(\mu} R_{\nu)\la}
 + \Box R_{\mu\nu} + \frac12 (\Box R -R_{\la\rho}^2 )g_{\mu\nu},
\eea
Let us observe that in the case $Z_N=0$ and $d=4$
the Einstein condition
\be
\label{einstein}
R_{\mu\nu}=\frac{1}{d}R g_{\mu\nu} ,
\ee
is enough to fulfil the equations of motion.
The scalar curvature remains an undetermined constant,
since the action is scale-invariant.
On the other hand when $Z_N\not=0$ the equations of motion
together with the Einstein condition further require
\be
\mathcal{P}\equiv
\frac{4-d}{2}\left(\alpha+\frac{\beta}{d}\right)R^2
\mp\frac{d-2}{2}Z_N\Big( R -\frac{2d}{d-2} \Lambda\Big) =0 .
\label{geneom}
\ee
In particular, in four dimensions the higher-derivative terms
do not contribute and the equation of motion is the same
as the trace of the Einstein equations with cosmological constant:
\be
\label{eom}
E\equiv R-\frac{2d}{d-2}\Lambda=0 \ .
\ee

To summarize, the choice of the Einstein condition
means that the background is ``almost on shell''
in the sense that all the equations of motion are satisfied
except, in general, for the trace equation (\ref{geneom}).
In the special case of pure HDG (no Einstein-Hilbert term) in $d=4$
the background is completely on-shell.

\subsection{Quadratic expansion}

The action has to be thought of as a functional of the
quantum field $\gamma_{\mu\nu}$ or $\gamma^{\mu\nu}$
defined as in (\ref{denslin}) or (\ref{densinv}).
For $w\not=-1/d$ 
these relations can be inverted to yield
\begin{equation}
g_{\mu\nu}=\gamma_{\mu\nu}\left(\det(\gamma_{\mu\nu})\right)^{m}
\ ;\qquad
g^{\mu\nu}=\gamma^{\mu\nu}\left(\det(\gamma_{\mu\nu})\right)^{-m}\ ,
\label{fc2}
\end{equation}
where 
\be
\frac{w}{2}=-\frac{m}{1+dm}\qquad
\mathrm{or}\qquad
\frac{w}{2}=\frac{m}{1+dm}\ ,
\label{relation}
\ee
respectively.
Conversely, $m=-\frac{w/2}{1+dw/2}$ for (\ref{denslin})
and $m=\frac{w/2}{1-dw/2}$ for (\ref{densinv}).
We observe that the relation between $m$ and $w/2$ is an involution.
We choose to treat $m$ as an independent free parameter.
All dependence on $m$ can be translated into a dependence on $w$
if needed, using the preceding formulas.

The quantum field is then expanded as in
(\ref{linexp}) or (\ref{expexp}) or (\ref{invexp}),
and for the calculation of the one-loop divergences we 
need the terms to second order in the fluctuation.
As explained in \cite{I}, we can start from the quadratic expansion
of the action in $\delta g_{\mu\nu}$, which has been given in
detail in \cite{OP}, and then use
\be
\label{gammaexp}
\delta g_{\mu\nu}=
\delta g^{(1)}_{\mu\nu}
+\delta g^{(2)}_{\mu\nu}
+\ldots\ ,
\ee
where $\delta g^{(n)}_{\mu\nu}$ contains $n$ powers 
of the tensor fluctuation
\be
\label{redefh}
h_{\mu\nu}=(\det\bar\gamma)^m\hat h_{\mu\nu}\ .
\ee
For all four types of expansion considered here, we have
\bea
\delta g^{(1)}_{\mu\nu}&=&h_{\mu\nu}+m\bg_{\mu\nu}h\ ,
\nonumber\\
\delta g^{(2)}_{\mu\nu}&=&
\omega h_{\mu\rho}h^\rho{}_\nu
+m h h_{\mu\nu}
+m\left(\omega-\frac{1}{2}\right)\bg_{\mu\nu}h^{\alpha\beta}h_{\alpha\beta}
+\frac{1}{2}m^2\bg_{\mu\nu}h^2\ .
\label{deltag}
\eea
The choice $\omega=0$ corresponds to the
linear expansion of metric (\ref{linexp}),
$\omega=1/2$ corresponds to the exponential expansion (\ref{expexp})
(and it does not matter if one starts from the metric
or from the inverse metric) and $\omega=1$ corresponds to the 
linear expansion of the inverse metric (\ref{invexp}).

In the following we will use these expansions
in the action (\ref{action}).
We note that for given $g_{\mu\nu}$ and $\bg_{\mu\nu}$,
different values of $\omega$ and $m$ will give different
fluctuation fields $h_{\mu\nu}$ and conversely
for given $h_{\mu\nu}$ and $\bg_{\mu\nu}$,
different values of $\omega$ and $m$ will give different
total metrics $g_{\mu\nu}$.
In the following calculations the action (\ref{action}),
as a functional of the total metric, will always be kept fixed
and also the functional measure for the quantum field
will be kept fixed.

\subsection{Lichnerowicz Laplacians}

The evaluation of the one-loop divergences is based on the
knowledge of the coefficients of the small-time expansion
of the heat kernel of the kinetic operator appearing in the
gauge-fixed Hessian.
We will denote $\bnabla$ the covariant derivative defined 
by the background metric $\bar{g}$,
$\Box=\bnabla^2$ the d'Alembertian and
$-\bnabla^2$ its Euclidean analog, known as Bochner Laplacian.
In GR, the expansion of the action generates non-minimal terms that
can be eliminated by choosing the de Donder gauge.
The rest can be written in terms of Laplacians.
Similarly, the second variation of the HDG action 
contains non-Laplacian terms such as 
$\bnabla_\mu\bnabla_\nu\bnabla_\rho\bnabla_\sigma$,
$\Box\bnabla_\mu\bnabla_\nu\bg_{\rho\sigma}
+\bg_{\mu\nu}\bnabla_\rho\bnabla_\sigma\Box$ and
$\bg_{\nu\sigma}\bnabla_\mu\Box\bnabla_\rho+
\bg_{\mu\rho}\bnabla_\nu\Box\bnabla_\sigma$,
acting on $h_{\rho\sigma}$.  
The heat kernel expansion for such operators is not known,
so we will use the York decomposition to rewrite the Hessian
as minimal operators acting on fields with definite spin.
For this, it is essential to rewrite all occurrences
of $\Box$ in terms of the Lichnerowicz Laplacians
acting on spin-zero, spin-one and spin-two fields,
which are defined (in Euclidean signature) as follows:
\bea
\lich_0 \phi &=& -\bnabla^2 \phi, \nn
\lich_1 A_\mu &=& -\bnabla^2 A_\mu + \bR_\mu{}^\rho A_\rho, \nn
\lich_2 h_{\mu\nu} &=& -\bnabla^2 h_{\mu\nu} 
+\bR_\mu{}^\rho h_{\rho\nu}
+ \bR_\nu{}^\rho h_{\mu\rho} 
-\bR_{\mu\rho\nu\s} h^{\rho\s} 
-\bR_{\mu\rho\nu\s} h^{\s\rho}\ .
\eea
These operators have the following useful properties:
\bea
\label{llprop1}
\lich_1\bnabla_\mu\phi&=&\bnabla_\mu\lich_0\phi,
\\
\label{llprop2}
\bnabla_\mu\lich_1\xi^\mu&=&\lich_0\bnabla_\mu\xi^\mu,
\\
\label{llprop3}
\lich_2 (\bnabla_\mu \bnabla_\nu \phi)
&=& \bnabla_\mu \bnabla_\nu \lich_0\phi,
\\
\label{llprop4}
\lich_2(\bnabla_\mu \xi_\nu + \bnabla_\nu \xi_\mu)
&=& \bnabla_\mu \lich_1 \xi_\nu + \bnabla_\nu \lich_1 \xi_\mu,
\\
\label{llprop5}
\lich_2 \bg_{\mu\nu}\phi&=&\bg_{\mu\nu}\lich_0\phi .
\eea
The York decomposition leads to significant simplifications
when the background metric is an Einstein space,
i.e. satisfies the condition (\ref{einstein}).  
(As discussed in section 2.1,
this is not enough to put the background on-shell.)
In the following we will always assume that the background is Einstein,
but not that it satisfies (\ref{eom}).

Now consider the second variations of the curvature squared 
terms in the action, as given in eqs.~(B.6) and (B.7) in \cite{OP}.
The terms containing the Riemann tensor can be combined with certain terms containing $\Box$ to give 
Lichnerowicz Laplacians.
Further using (\ref{einstein}), one arrives at
\bea
&&
\a \delta g^{\mu\nu} \Big[ \bnabla_\mu \bnabla_\nu \bnabla_\a \bnabla_\b
- 2\bg_{\mu\nu} \Box \bnabla_\a \bnabla_\b 
+\bg_{\mu\nu} \bg_{\a\b} \Box^2
\nn
&&
-\bg_{\nu\b}\br\bnabla_\mu \bnabla_\a
+\frac{d-2}{2}\br\bg_{\mu\nu}\bnabla_\a \bnabla_\b
+\frac{4-d}{2d}\bg_{\mu\nu}\bg_{\a\b}\br\Box
-\frac12\bg_{\mu\a}\bg_{\nu\b}\br\lich_2
\nn
&&
+\left(\frac{1}{d^2}-\frac{1}{d}+\frac{1}{8}\right)
\br^2\bg_{\mu\nu} \bg_{\a\b}
+\left(\frac{2}{d}-\frac{1}{4}\right)
\br^2\bg_{\mu\a} \bg_{\nu\b}
\Big] \delta g^{\a\b},
\label{varalpha2}
\eea
and
\bea
\label{varbeta2}
\beta \delta g^{\mu\nu}\hs{-7}&&
\Big[\frac{1}{2}\bnabla_\mu \bnabla_\nu \bnabla_\a \bnabla_\b
-\frac{1}{2}\bg_{\mu\nu} \Box \bnabla_\a \bnabla_\b
-\frac{1}{2}\bg_{\nu\b} \bnabla_\mu \Box \bnabla_\a
+\frac14\bg_{\mu\nu} \bg_{\a\b}\Box^2
\nonumber
\\
&&
+\frac14\bg_{\mu\a}\bg_{\nu\b}\lich_2\left(\lich_2-\frac{6}{d}\br\right)
-\frac{3}{2d}\br\bg_{\nu\b} \bnabla_\mu \bnabla_\a
+\frac{1}{d}\br \bg_{\mu\nu} \bnabla_\a\bnabla_\b
-\frac{1}{4d}\br\bg_{\mu\nu}\bg_{\a\b} \Box
\nonumber
\nn
&&
+\frac{12-d}{4d^2}\br^2\bg_{\mu\a} \bg_{\nu\b}
+\frac{d-8}{8d^2}\br^2\bg_{\mu\nu} \bg_{\a\b}
\Big] \delta g^{\a\b}.
\eea
All tensor structures are provided by the background metric.
We note that this procedure of eliminating the Riemann tensor
does not work in the case of the second variation of the Riemann squared term,
which is the reason why we do not consider such a term.

\subsection{York decomposition}

The York decomposition is defined by
\bea
h_{\mu\nu} = h^{TT}_{\mu\nu} 
+ \bnabla_\mu\xi_\nu 
+ \bnabla_\nu\xi_\mu
+\bnabla_\mu \bnabla_\nu \s 
-\frac{1}{d} \bg_{\mu\nu} \bnabla^2 \s +
\frac{1}{d} \bg_{\mu\nu} h,
\label{york}
\eea
where $h^{TT}_{\mu\nu}$ is transverse and tracefree,
and $\hat\xi_\mu$ is transverse.
We will use
\be
\hat{\xi}_{\mu}=\sqrt{-\bar{\nabla}^{2}-\frac{\bar{R}}{d}}\xi_{\mu}
\ \ ;\qquad
\hat{\sigma}=\sqrt{-\bar{\nabla}^{2}}\sqrt{-\bar{\nabla}^{2}-\frac{\bar{R}}{d-1}}\sigma\,.
\label{ap6}
\ee

Employing (\ref{york}) and (\ref{ap6}) and assuming that the background is an Einstein space, 
one finds
\be
\int d^dx\sqrt{\bg}h_{\mu\nu}h^{\mu\nu}=
\int d^d x\sqrt{\bg}\Big[h^{TT}_{\mu\nu}h^{TT\,\mu\nu} 
+2 \hat\xi_\mu \hat\xi^\mu 
+\frac{d-1}{d}\hat\s^2
+\frac{1}{d} h^2 \Big].
\ee
Using the properties (\ref{llprop1}) -- (\ref{llprop5}), we have
\be
\int d^dx\sqrt{\bg}h_{\mu\nu}\lich_2 h^{\mu\nu}=
\int d^d x\sqrt{\bg}\Big[ h^{TT}_{\mu\nu}\lich_2h^{TT\,\mu\nu} 
+2\hat\xi_\mu\lich_1\hat\xi^\mu 
+\frac{d-1}{d}\hat\s \lich_0\hat\s
+\frac{1}{d} h\lich_0 h \Big] ,
\ee
\bea
\int d^dx\sqrt{\bg}h_{\mu\nu}\left(\lich_2\right)^2 h^{\mu\nu}=
\int d^d x\sqrt{\bg}\Big[ h^{TT}_{\mu\nu}\left(\lich_2\right)^2h^{TT\,\mu\nu} 
+2\hat\xi_\mu\left(\lich_1\right)^2\hat\xi^\mu \nn
+\frac{d-1}{d}\hat\s \left(\lich_0\right)^2\hat\s
+\frac{1}{d} h\left(\lich_0\right)^2 h \Big] ,
\eea
and so on.

Furthermore using
\be
\bnabla_\a h^\a{}_\nu=\left(\bnabla^2+\frac{\br}{d}\right)\xi_\nu
+\frac{d-1}{d}\bnabla_\nu\left(\bnabla^2+\frac{\br}{d-1}\right)\s
+\frac{1}{d}\bnabla_\nu h\ ,
\ee
and
\be
\bnabla_\a \bnabla_\b h^{\a\b}=
\frac{d-1}{d}\bnabla^2\left(\bnabla^2+\frac{\br}{d-1}\right)\s
+\frac{1}{d}\bnabla^2 h\ ,
\ee
we find that
\bea
\int d^dx\sqrt{\bg}
h^{\mu\nu}\bnabla_\mu\bnabla_\nu\bnabla_\a\bnabla_\b h^{\a\b}
&=&
\int d^dx\sqrt{\bg}\Bigg[
\left(\frac{d-1}{d}\right)^2
\s\Box^2\left(\Box+\frac{\bR}{d-1}\right)\s
\nonumber\\
&&+2\frac{d-1}{d^2}h\Box^2\left(\Box+\frac{\bR}{d-1}\right)\s
+\frac{1}{d^2}h\Box^2 h
\Bigg],
\eea
\bea
\int d^dx\sqrt{\bg}
h^{\mu\nu}\bnabla_\mu\bnabla_\a {h^{\a}}_{\nu}
&=&
\int d^dx\sqrt{\bg}\Bigg[
-\xi_\mu\left(\Box+\frac{\br}{d}\right)^2\xi^\mu
+\left(\frac{d-1}{d}\right)^2
\s\Box\left(\Box+\frac{\bR}{d-1}\right)^2\s
\nonumber\\
&&+2\frac{d-1}{d^2}h\Box\left(\Box+\frac{\bR}{d-1}\right)\s
+\frac{1}{d^2}h\Box h
\Bigg],
\eea
\bea
\int d^dx\sqrt{\bg}
h^{\mu\nu}\bnabla_\mu\Box\bnabla_\a {h^{\a}}_{\nu}
&=&
\int d^dx\sqrt{\bg}\Bigg[
-\xi_\mu\Box\left(\Box+\frac{\br}{d}\right)^2\xi^\mu
\nonumber\\
&&
+\left(\frac{d-1}{d}\right)^2
\s\Box\left(\Box+\frac{\br}{d}\right)\left(\Box+\frac{\bR}{d-1}\right)^2\s
\nonumber\\
&&+2\frac{d-1}{d^2}h\Box\left(\Box+\frac{\br}{d}\right)\left(\Box+\frac{\bR}{d-1}\right)\s
+\frac{1}{d^2}h\Box\left(\Box+\frac{\br}{d}\right) h\Bigg].\nn
\eea

\subsection{The decomposed Hessian}
\label{decomposed}

The expansion of the Euclidean action in powers of $h$
has the following quadratic part
\be
S^{(2)}=
\int d^dx\sqrt{\bg}
\left[
h^{TT}_{\mu\nu}H^{TT}h^{TT\mu\nu}
+\hat\xi_\mu H^{\xi\xi}\hat\xi^\mu
+\hat\sigma H^{\sigma\sigma}\hat\sigma
+\hat\sigma H^{\sigma h} h
+h H^{h\sigma} \hat\sigma
+h H^{hh} h
\right],
\ee
with
\bea
H^{TT}&=&
\frac{1}{4}Z_N\left(\lich_2-\frac{2\bR}{d}
+\frac{d-2}{d}(1-2\omega)(1+dm)\eom
\right)
\nonumber\\
&&
+\frac{\beta}{4}\left(\left(\lich_2\right)^2
-\frac{6}{d}\bR\lich_2
+\frac{8-(d-4)(1-2\omega)(1+dm)}{d^2}\bR^2\right)
\nonumber\\
&&
-\frac{\alpha}{2}\bR\left(
\lich_2-\frac{4-(d-4)(1-2\omega)(1+dm)}{2d}\bR\right),
\label{hTTTT}
\eea
\bea
H^{\xi\xi}&=&
\frac{(1-2\omega)(1+dm)}{2d}\left[
(d-2)Z_N\eom
-\frac{d-4}{d}(d\alpha+\beta)\bR^2\right],
\eea
\bea
H^{\sigma\sigma}&=&
\frac{d-1}{d}\Bigg\{
-\frac{d-2}{4d}Z_N
\left(\lich_0-(1-2\omega)(1+dm)\eom\right)
\nonumber\\
&&+\frac{d-1}{d}\alpha\left[\left(\lich_0\right)^2
+\frac{d-4}{2(d-1)}\bR
\left(\lich_0-\frac{(1-2\omega)(1+dm)}{2}\bR\right)\right]
\nonumber\\
&&+\frac{\beta}{4}\left[\left(\lich_0\right)^2
+\frac{2(d-4)}{d^2}\bR
\left(\lich_0-\frac{(1-2\omega)(1+dm)}{2}\bR\right)\right]
\Bigg\},
\label{hsigsig}
\eea
\bea
H^{\sigma h}=H^{h\sigma}&=&\frac{d-1}{2d}
(1+dm)\sqrt{\lich_0}\sqrt{\lich_0-\frac{\bR}{d-1}}
\Bigg[-Z_N\frac{d-2}{2d}
\nonumber\\
&&
+\alpha\frac{2(d-1)}{d}\left(\lich_0+\frac{d-4}{2(d-1)}\bR\right)
+\frac{\beta}{2}\left(\lich_0+\frac{2(d-4)}{d^2}\bR\right)\Bigg],
\eea
\bea
H^{hh}&=&\frac{d-1}{d}(1+dm)^2\Bigg\{
-Z_N\frac{d-2}{4d}
\left(\lich_0-\frac{\bR}{d-1}
+\frac{d-2+d^2 m+4\omega}{2(d-1)(1+dm)}\eom\right)
\nonumber\\
&&+\alpha\frac{d-1}{d}\left(\left(\lich_0\right)^2
+\frac{d-6}{2(d-1)}\bR\,\lich_0
+\frac{(d-4)(d-6-4 d m + d^2 m + 4 \omega)}{8(d-1)^2(1+dm)}\bR^2
\right)
\nonumber\\
&&+\frac{\beta}{4}\left(\left(\lich_0\right)^2
+\frac{d^2-10d+8}{d^2(d-1)}\bR\,\lich_0
+\frac{(d-4)(d-6-4 d m + d^2 m + 4 \omega)}{2(d-1)d^2(1+dm)}\bR^2
\right)
\Bigg\},\nonumber\\
\eea
where $\eom=\bR-\frac{2d\Lambda}{d-2}$.

If we define the scalar gauge-invariant degree of freedom
\be
s=\frac{\sqrt{\lich_0}}{\sqrt{\lich_0-\frac{\bR}{d-1}}}\hat\sigma
+(1+dm)h,
\label{invscalar}
\ee
the scalar sector of the Hessian can be rewritten as
\be
\int d^dx\sqrt{\bg}
\left[
s H_s^{ss}s
+s H_s^{sh} h
+h H_s^{hs} s
+h H_s^{hh} h
\right],
\ee
where
\bea
H_s^{ss}&=&
\frac{d-1}{4d^3}\frac{\lich_0-\frac{\bR}{d-1}}{\lich_0}
\Bigg\{
-d(d-2)Z_N\left[\lich_0
-(1+dm)(1-2\omega)\left(\bR-\frac{2d\Lambda}{d-2}\right)\right]
\nonumber\\
&&+4d(d-1)\alpha\left[\left(\lich_0\right)^2
+\frac{d-4}{2(d-1)}\lich_0 \bR
-\frac{d-4}{4(d-1)}(1+dm)(1-2\omega)\bR^2\right]
\nonumber\\
&&
+d^2\beta\left[\left(\lich_0\right)^2
+2\frac{d-4}{d^2}\lich_0 \bR
-\frac{d-4}{d^2}(1+dm)(1-2\omega)\bR^2\right]
\Bigg\},
\label{hsss}
\\
H_s^{s h}&=&H_s^{h s}=
-\frac{d-1}{2d^2}\frac{\lich_0-\frac{\bR}{d-1}}{\lich_0}
(1+dm)^2(1-2\omega)\bar{\mathcal{P}},
\label{hssh}
\\
H_s^{hh}&=&-\frac{1}{4d^2}\frac{(1+dm)}{\lich_0}
\Big[d\Big( 2d(d-1)m^2(2\omega-1)+dm (8\omega-3)-4m(2\omega-1)+4\omega -1\Big)
\lich_0
\nonumber\\
&&
+2(1+dm)^2(1-2\omega)\bR\Big]
\label{hshh}
\bar{\mathcal{P}}.
\eea

This form of the Hessian has the virtue that all the terms
that contain $h$ are proportional to the equation of motion.
This shows that the field $h$ can be completely disregarded on shell,
as one would expect of a gauge-variant variable.
This however depends on the choice of basis in the space of fields
and is only true when the other scalar degree of freedom 
is gauge-invariant.

On the other hand, this form of the Hessian has the 
unpleasant feature that 
the kinetic operator of the field $s$ is non-local.
One cannot generally absorb the non-local prefactor in
a redefinition of $s$, because one is not allowed to
perform non-local redefinitions of physical fields.

One notices, however, that the terms with the lowest power
of $\lich_0$ in each of the three lines in $H_s^{ss}$ 
is proportional to $(1+dm)(1-2\omega)$.
Therefore, if either $\omega=1/2$
(exponential parametrization)
or $m=-1/d$ (the ``unimodular'' measure),
each of the square brackets in (\ref{hsss})
is proportional to $\lich_0$ and 
the Hessian of $s$ becomes local.

We further observe that for $\omega=1/2$
also the mixed term vanishes and the term $H_s^{hh}$ becomes local,
whereas for $m=-1/d$ all terms containing $h$ vanish,
as expected in the unimodular theory.

Furthermore, $H_s^{ss}$ becomes local 
and independent of $m$ and $\omega$ for pure 
four-derivative gravity ($Z_N=0$) when $d=4$.
In this case the whole Hessian becomes just
\bea
\label{hdgh}
\frac14 \b h_{\mu\nu}^{TT} \Big(\Delta_{L2}-\frac{\br}{2}\Big)
\Big(\Delta_{L2}-\Big(1+\frac{2\a}{\b}\Big)\br\Big)h^{TT\mu\nu}
+\frac{3}{16}(3\a+\b)s\Delta_{L0} \Big(\Delta_{L0}-\frac{\br}{3}\Big)s.
\eea
Finally we observe that in the conformal case $d=4$, 
$\beta=-3\alpha$
(which corresponds to having only the term $W$ in the action)
the $ss$ term in the Hessian vanishes too, leaving only the
pure spin-two degree of freedom, with Hessian
\bea
\label{hdghc}
\frac14 \b h_{\mu\nu}^{TT} \left(\Delta_{L2}-\frac{\br}{2}\right)
\left(\Delta_{L2}-\frac{\br}{3}\right)h^{TT\mu\nu}\ .
\eea

\subsection{Two-derivative gauge fixing terms}

We now turn to the discussion of the gauge-fixing and Faddeev-Popov (FP) ghost terms.
In most of the following, we use the same two-derivative gauge-fixing term as in I \cite{I}.
The gauge-fixing function is defined as
\begin{equation}
F_{\mu}=\bar{\nabla}_{\alpha}{h^{\alpha}}_{\mu}-\frac{\bar b+1}{d}\bar{\nabla}_{\mu}h\,,
\label{gaugecondition}
\end{equation}
with a gauge-fixing parameter $-\infty<\bar b<\infty$.
Here $h_{\mu\nu}$ is the tensorial quantum field defined above
and the bars on the covariant derivatives means that they are calculated
from the background metric $\bg_{\mu\nu}$.
The simplest way to derive the gauge-fixing and FP ghost terms is to use the BRST transformations
which is obtained by replacing the gauge parameters with the FP ghosts. In the present case, we have
\bea
\d_B g_{\mu\nu}&=& -\d \la [ g_{\rho\nu}\nabla_\mu C^\rho
 + g_{\rho\mu}\nabla_\nu C^\rho ] ,
\eea
where $C^\mu$ is the FP ghost, and $\d\la$ is an anticommuting parameter.
The BRST transformation for other fields is derived by the requirement of the nilpotency of
the transformation:
\bea
\d_B C^\mu = \d\la C^\rho \pa_\rho C^\mu,~~~
\d_B \bar C_\mu = i \d\la\, B_\mu, ~~~
\d_B B_\mu = 0,
\label{brst}
\eea
where $\bar C_\mu$ is the FP anti-ghost and $B_\mu$ is an auxiliary field which enforces the gauge-fixing condition.

In order to derive the gauge fixing and FP terms, we have to know the transformations
on the fluctuation field $h_{\mu\nu}$. From the first relation in eq.~\p{deltag}, we have
\bea
\d_B h_{\mu\nu} + m \bg_{\mu\nu} \d_B h + \cdots
= -\d \la [ g_{\rho\nu}\nabla_\mu C^\rho + g_{\rho\mu}\nabla_\nu C^\rho ].
\eea
Anticipating that we will compute the effective action for vanishing expectation value of $h_{\mu\nu}$,
we can restrict this to the linear terms to obtain
\bea
\d_B h_{\mu\nu}
= -\d \la \left[ \bnabla_\mu C_\nu + \bnabla_\nu C_\mu
- \frac{2m}{1+dm} \bg_{\mu\nu} \bnabla_\rho C^\rho \right].
\eea
The gauge-fixing and FP terms are then given as~\cite{HKO}
\bea
{\cal L}_{GF+FP}/\sqrt{\bg} \hs{-2}&=&\hs{-2}
i \d_B \left[\bar C_\mu \left(F^\mu+\frac{a}{2Z_{GF}} B^\mu\right)\right]/\d\la \nn
\hs{-2} &=&\hs{-2} - B_\mu \left( F^\mu +\frac{a}{2Z_{GF}} B^\mu\right)
+ i \bar C_\mu \left[ \bnabla_\a \left( \bnabla^\mu C^\a+\bnabla^\a C^\mu
-\frac{2m}{1+dm} \bg^{\mu\a}\bnabla_\rho C^\rho\right) \right. \nn
&& \hs{30} - \left. \frac{1+\bar b}{d} \bnabla^\mu\left(\frac{2}{1+dm} \bnabla_\rho C^\rho\right) \right]
  \nn
&=&\hs{-2} -\frac{a}{2Z_{GF}} \tilde B_\mu \tilde B^\mu + \frac{Z_{GF}}{2a} F_\mu F^\mu
+ i \bar C_\mu \Delta^{(gh)\mu}{}_{\nu} C^\nu,
\label{gfghl}
\eea
and we have set $\bar b=(1+dm)b$ and defined
\bea
\tilde B_{\mu} = B_\mu +\frac{Z_{GF}}{a} F_\mu, \qquad
\Delta^{(gh)}_{\mu\nu} \equiv \bg_{\mu\nu} \Box+
\Big(1-2 \frac{b+1}{d}\Big) \bnabla_\mu \bnabla_\nu +\br_{\mu\nu}.
\label{bfield}
\eea
Here $a$ is another dimensionless gauge parameter, and
$\zgf$ is a parameter with dimension $d-2$.
Since the $B_\mu$ field involves no derivatives, we can simply integrate it out and then we are left with
the gauge-fixing and FP ghost terms.

Using the York decomposition,
\be
F_\mu=-\left(\lich_1-\frac{2\bR}{d}\right)\xi_\mu
-\bnabla_\mu
\left(\frac{d-1}{d}\left(\lich_0-\frac{\bR}{d-1}\right)\sigma
+\frac{\bar b}{d} h\right) \ .
\ee
Inserting this into the gauge-fixing term in (\ref{gfghl}), integrating by parts,
rewriting in terms of Lichnerowicz Laplacians 
and using the York decomposition, we obtain the gauge-fixing term
\bea
S_{GF}&=&-\frac{\zgf}{2a}\int d^dx\,\sqrt{\bg}
\Bigg[\hat\xi_\mu\left(\lich_1-\frac{2\bR}{d}\right)\hat\xi^\mu
+\frac{(d-1)^2}{d^2}\hat\sigma
\left(\lich_0-\frac{\bR}{d-1}\right)\hat\sigma
\nonumber
\\
&&\qquad\qquad\qquad+\frac{(d-1)\bar b}{d^2}\hat\sigma
\sqrt{\lich_0}\sqrt{\lich_0-\frac{\bR}{d-1}}h
+\frac{\bar b^2}{d^2}h\lich_0 h
\Bigg]\,.
\label{gfa2}
\eea
We see that a specific combination of scalar degrees of freedom 
appears in the gauge-fixing term.
It is sometimes convenient to write the scalar sector in terms of the
gauge-invariant variable $s$ defined in eq.~\p{invscalar} and this new degree of freedom which,
in terms of the original fields, is given by
\be
\label{chi}
\chi=\sigma+\frac{b}{(d-1-b)\lich_0-\bR}s 
= \frac{(d-1)\lich_0-\bR}{(d-1-b)\lich_0-\bR}\sigma
+\frac{b(1+dm)}{(d-1-b)\lich_0-\bR}h\ .
\label{newf}
\ee
A short calculation shows that the Jacobian of the transformation $(\s,h)\to(s,\chi)$
is one.

In terms of the new variable, the gauge-fixing action is
\be
\label{gf2}
S_{GF}= - \frac{\zgf}{2a}\int d^dx\sqrt{\bg}
\left[
\xi_\mu\left(\lich_1-\frac{2\bR}{d}\right)^2\xi^\mu
+\frac{(d-1-b)^2}{d^2}
\chi\; \lich_0\left(\lich_0-\frac{\bR}{d-1-b}\right)^2\chi
\right].
\ee

From (\ref{chi}) we see that $\chi$ transforms in the same way as $\sigma$.
Thus $\xi$ and $\chi$ can be viewed as the gauge degrees of freedom and
$h^{TT}$ and $s$ as the physical degrees of freedom.
We will use this parametrization later.

The ghost action~\p{gfghl} for this gauge-fixing contains a non-minimal operator~\cite{I}.
We also decompose the ghost into transverse and longitudinal parts
\be
C_\nu=C^T_\nu+\bnabla_\nu C^L
=C^T_\nu+\bnabla_\nu\frac{1}{\sqrt{-\bnabla^2}}C'^L,
\ee
and the same for $\bar C_\mu$.
(This change of variables has unit Jacobian).
The ghost action then splits in two terms
\be
\label{ghostaction}
S_{gh}= i \int d^dx\sqrt{\bg}
\left[
\bar C^{T\mu}\left( \lich_1-\frac{2\bR}{d}\right)C^T_\mu
+2\frac{d-1-b}{d}
\bar C'^L\left(\lich_0-\frac{\bR}{d-1-b}\right)C'^L\right].
\ee
Note for future reference that the gauge-fixing condition 
and the ghost action are pathological for $b=d-1$.

Let us observe that $\zgf$ appears in the
combination $\zgf/a$, where $a$ is a dimensionless 
gauge parameter, while $\zgf$ is a constant of dimension $d-2$. 
%
%
%
There are then two natural options regarding the constant $\zgf$.
The first choice is to treat it as a fixed parameter
(later we shall identify it with a power of the cutoff).
This leads to simpler formulas for many expressions,
and we shall use it extensively later.
It is however not appropriate for the discussion of perturbative
Einstein gravity.
The reason is that in the limit $G\to0$ 
the coefficient $Z_N$ of the Hessian diverges.
If we keep $\zgf$ and $a$ constant in the limit,
then the gauge-fixing term becomes negligible
relative to the rest of the quadratic action.
The gauge fluctuations remain unsuppressed and one can anticipate divergences.
This is exactly what happens, as we shall mention later on.
One can compensate the behavior of $Z_N$ by keeping $\zgf$ fixed
and letting simultaneously $a\to0$.
Alternatively, one can set $\zgf=Z_N$.
In this case, in the Gaussian limit $G\rightarrow 0$,
the kinetic terms of the gauge-invariant and gauge degrees of
freedom scale in the same way and one obtains 
sensible results for all values of $a$ and $b$.

In the following we shall discuss also different choices
for the gauge fixing. One is the so-called 
``unimodular physical gauge'', where one sets
$\hat\xi_\mu=0$ and $h=0$.
As shown in \cite{vacca}, this is equivalent 
to the above standard
gauge in the limit $b \to\pm\infty$ and $a\to0$.

\subsection{Four-derivative gauge-fixing terms}

In HDG it is customary to use gauge-fixing terms that
contain four derivatives.
In order to further check gauge-independence of the results,
in section \ref{quarticg}
we consider the gauge fixing in our previous paper~\cite{OP} but now with arbitrary gauge parameters.

We can simply repeat the same procedure to derive the gauge-fixing and FP terms but with additional
higher derivative operator
\bea
Y_{\mu\nu} \equiv \bg_{\mu\nu} \Box+ c \bnabla_\mu \bnabla_\nu 
- f \bnabla_\nu \bnabla_\mu,
\eea
where $c$ and $f$ are additional gauge parameters:
\bea
{\cal L}_{GF+FP}/\sqrt{\bg} \hs{-2}&=&\hs{-2}
i \d_B \left[\bar C_\mu Y^{\mu\nu}\left(F_\nu-\frac{a}{2\zgf} B_\nu\right)\right]/\d\la \nn
&=&\hs{-2} -\frac{a}{2\zgf} \tilde B_\mu Y^{\mu\nu} \tilde B_\nu +\frac{\zgf}{2a} F_\mu Y^{\mu\nu} F_\nu
+ i \bar C_\mu Y^{\mu\nu} \Delta^{(gh)}_{\nu\rho} C^\rho,
\label{gfgh}
\eea
where the gauge-fixing function $F_\mu$ is defined in \p{gaugecondition}, and
$\tilde B_\mu$ and $\Delta^{(gh)}_{\mu\nu}$ in \p{bfield}.
Here we should notice that the dimension of the constant $\zgf$ is changed to $d-4$.

Note also that the field $B_\mu$ was an auxiliary field in the two-derivative gauge-fixing case~\p{gfghl},
but here it is dynamical. With higher derivative gauge fixing, quite often only the contribution
$\Delta^{(gh)}_{\mu\nu}$ is incorporated but that from $Y^{\mu\nu}$ in the FP ghost kinetic term
is ignored, and then it is claimed that somehow the contribution
from the ``third ghost'' $-\frac12 \log(\det(Y_{\mu\nu}))$ must be added.
We see here that this is automatic in the BRST invariant formulation, because we have contributions
$-\log(\det(Y^{\mu\nu}))$ from the FP ghost kinetic term and $\frac12 \log(\det(Y^{\mu\nu}))$ from
the field $B_\mu$, giving the same result.

Using the York decomposition~\p{york}, we can calculate the contributions without
fixing gauge parameters and so can check the gauge dependence directly here.
We find that the terms~\p{gfgh} are cast into
\bea
&&
- \frac{a}{2\zgf} \Big[ B_\mu^T \Big(\Delta_{L1}-\frac{1-f}{d}\br \Big) B^{\mu T}
+B^L \Delta_{L0}\Big\{(1+c-f)\Delta_{L0}-\frac{1-f}{d}\br \Big\} B^L \Big] \nn
&&
+\frac{\zgf}{2a} \Big[ \xi_\mu \Big( \Delta_{L1}-\frac{2}{d} \br \Big)^2
\Big(\Delta_{L1}-\frac{1-f}{d}\br \Big) \xi^\mu \nn
&& \hs{15}
+\Big(\frac{d-1-b}{d}\Big)^2 \chi \Delta_{L0}\Big(\Delta_{L0}-\frac{\br}{d-1-b}\Big)^2
\Big\{(1+c-f)\Delta_{L0}-\frac{1-f}{d}\br\Big\} \chi \Big] \nn
&& +i \bar C_\mu^T \Big(\Delta_{L1}-\frac{1-f}{d} \br \Big)\Big(\Delta_{L1}-\frac{2}{d}\br \Big)
C^{\mu T} \nn
&&
+2i \frac{d-1-b}{d} \bar C^L \Delta_{L0}\Big(\Delta_{L0}-\frac{\br}{d-1-b} \Big)
\Big\{(1+c-f)\Delta_{L0}-\frac{1-f}{d} \br \Big\} C^{L},
\eea
where we have defined transverse and longitudinal parts of the $\tilde B_\mu$ and
FP ghosts $\bar C_\mu$ and $C_\mu$, as usual.

In section \ref{quadraticg} we will discuss the gauge-independence of
the theory in the general quadratic gauge.
In section 4.5 we will restrict ourselves to the 
class of gauges where $c=f=1$ but with generic $a$, $b$.
This is equivalent to inserting a Lichnerowicz Laplacian $\lich_1$
in the quadratic gauge fixing term (\ref{gfghl}),
or in other words to set $Y^{\mu\nu}=\bg^{\mu\nu}\lich_1$.
After performing the York decomposition, this yields
additional factors $\lich_1$ and $\lich_0$ in the quadratic
actions of $\xi_\mu$ and $\chi$, equation (\ref{gf2}).
The resulting additional determinants are offset by the
determinant of the operator coming from the $B_\mu$ sector, as will become
clear in the following.

\section{Universality on-shell in $d=4$}

In this section we consider the theory 
in $d=4$ on an Einstein background.
As noted in section 2.1, if we put $Z_N=0$, 
the equation of motion of HDG is automatically satisfied for $d=4$,
so one would expect the effective action to be gauge-
and parametrization-independent.
We will check that this is indeed the case at the formal level
of determinants.
In the following section we will have a more explicit check
of this property in the expressions for the divergences.

\subsection{General two-derivative gauge fixing}
\label{quadraticg}

The Hessian of pure higher-derivative gravity in $d=4$
has the simple form (\ref{hdgh}), independent of the 
choice of parametrization.
It is expressed entirely in terms of the gauge-invariant
variables $h^{TT}$ and $s$.
These fields do not appear in the gauge-fixing action,
so their contribution to the one-loop effective action
is given by:
\bea
\Det \Big(\Delta_{L2}-\frac{\br}{2} \Big)^{-1/2}
\Det \Big(\Delta_{L2}-\Big(1+\frac{2\a}{\b} \Big) \br \Big)^{-1/2}
\Det \Delta_{L0}^{-1/2}
\Det \Big(\Delta_{L0}-\frac{\br}{3} \Big)^{-1/2}\, .
\label{dets1}
\eea
We choose the gauge as in section 2.6. 
It is convenient to express the gauge fixing term 
as in (\ref{gf2}).
The fields $\xi$ and $\chi$ 
contribute to the one-loop action the terms (as ordered):
\bea
\Det \Big(\Delta_{L1}-\frac{\br}{2}\Big)^{-1}\,
\Det \Delta_{L0}^{-1/2}\,
\Det \Big(\Delta_{L0}-\frac{\br}{3-b} \Big)^{-1}\ .
\eea
The ghost action (\ref{ghostaction}) gives the determinants
\bea
\Det \Big(\Delta_{L1}-\frac{\br}{2} \Big)
\Det \Big(\Delta_{L0}-\frac{\br}{3-b} \Big)\, ,
\eea
and finally the Jacobians of the change of variables 
from $h_{\mu\nu}$ to $h^{TT}_{\mu\nu}$, $\xi^\mu$, $\sigma$ and $h$ is:
\bea
\Det \Big(\Delta_{L1}-\frac{\br}{2} \Big)^{1/2}\,
\Det \Delta_{L0}^{1/2}\,
\Det \Big(\Delta_{L0}-\frac{\br}{3} \Big)^{1/2}\ .
\label{jacs}
\eea
Note that the gauge parameter $a$ only appears in the Hessian in the prefactors and not in the
operators. The only dependence on the gauge parameter $b$ is in two determinants that cancel.
This proves the gauge independence of the one-loop action. The final result is
\bea
\frac
{\sqrt{\Det\left(\Delta_{L1}-\frac{\br}{2}\right)}}
{\sqrt{\Det\left(\Delta_{L2}-\frac{\br}{2}\right)}
\sqrt{\Det\left(\Delta_{L2}-\Big(1+\frac{2\a}{\b}\Big)\br \right)}
\sqrt{\Det\Delta_{L0}}}\ .
\label{eff1}
\eea

\subsection{General four-derivative gauge fixing}
\label{quarticg}

If we use the general four-derivative gauge fixing of section 2.7,
we have again the determinants (\ref{dets1})
coming from the physical degrees of freedom
and (\ref{jacs}) coming from the Jacobians.
The determinants coming from the fields $\xi_\mu$ and $\chi$ are
\bea
&&{\Det} \Big( \Delta_{L1}-\frac{\bR}{2}\Big)^{-1}
{\Det} \Big(\Delta_{L1}-\frac{1-f}{4}\br \Big)^{-1/2}\\
&&\hs{10}\times{\Det} \Big(\Delta_{L0}-\frac{\br}{3-b}\Big)^{-1}
{\Det} \Big(\Delta_{L0}-\frac{1-f}{4(1+c-f)}\br\Big)^{-1/2}\, ,
\eea
the ghosts give
\bea
&& {\Det} \Big(\Delta_{L1}-\frac{\bR}{2}\Big) 
{\Det} \Big(\Delta_{L1}-\frac{1-f}{4} \br \Big)\nn
&& \hs{10}\times\; {\Det} \Delta_{L0}\; {\Det} \Big(\Delta_{L0}-\frac{\br}{3-b} \Big)
{\Det} \Big(\Delta_{L0}-\frac{1-f}{4(1+c-f)} \br \Big),~~
\eea
and finally we have the contributions of the $B_\mu$ field
\bea
{\Det} \Big(\Delta_{L1}-\frac{1-f}{4}\br \Big)^{-1/2}
{\Det} \Delta_{L0}^{-1/2}~
{\Det} \Big(\Delta_{L0}-\frac{1-f}{4(1+c-f)}\br \Big)^{-1/2}\, .
\eea
Putting everything together, we see that all the terms 
depending on the gauge parameters cancel out
and the remaining ones give again (\ref{eff1}).
Thus we have explicitly shown, at this formal level, 
that the results are completely gauge independent.

\subsection{Physical gauge}

In the calculation of sections 3.1, and even more in section 3.2, 
there is a large number of cancellations 
between various determinants.
Consider instead the ``physical" gauge 
$\hat\xi_\mu=0$, $h=0$, discussed in \cite{ppps,vacca}. 
It leaves only the fields $h^{TT}$ and $\hat\sigma$, 
and no Jacobians. 
The Hessians of $h^{TT}$ and $\hat\sigma$ are 
given by (\ref{hTTTT}) and (\ref{hsigsig}) respectively.
There are a real scalar ghost and a real transverse vector ghost, 
with ghost operators $\Delta_{L0}$ and $\Delta_{L1}- R/2$.
Putting together these terms,
one immediately obtains the effective action \p{eff1}. 
In fact this is the most
direct way of getting it, because there is no cancellation of determinants between unphysical degrees of freedom and ghosts.

\subsection{The conformal case}

Now we consider the conformal case where $\beta/\alpha=-3$.
The effective action in this case cannot be simply 
obtained as a particular case of eq.(\ref{eff1}),
because the action is invariant also under Weyl transformations and
this requires a separate gauge fixing.

The Hessian is only nonzero in the spin-two sector
and is given by (\ref{hdghc}).
For the Weyl invariance we can gauge fix $h=0$,
without any ghost because $h$ transforms under Weyl transformations
by a shift.
For diffeomorphisms we choose a standard gauge fixing 
of the form (\ref{gfa2}).
Since $h=0$, the value of $b$ is immaterial.
Equation (\ref{chi}) is replaced by $\chi=\sigma$,
so the decomposition of the gauge-fixing and ghost actions,
and the corresponding determinants,
are the same as in section 2.6, with $b=0$.

We use the fields $\hat{\xi}$ and $\hat{\sigma}$ (see (\ref{ap6})) 
in such a way that no Jacobian is needed. 
The one-loop effective action is
\be
\label{confdet}
\frac
{\sqrt{\Det\left(\Delta_{L1}-\frac{\bR}{2}\right)}
\sqrt{\Det\left(\Delta_{L0}-\frac{\bR}{3}\right)}}
{\sqrt{\Det\left(\Delta_{L2}-\frac{\bR}{2}\right)}
\sqrt{\Det\left(\Delta_{L2}-\frac{\bR}{3}\right)}} \ .
\ee
The two terms in the denominator come from the TT part.
The terms in the numerator come from the ghosts (power $1$)
and from the gauge fixing term (power $-1/2$).
Note that there is no dependence on the gauge parameter $a$.
It only appears in the prefactors of the
quadratic action, which drop out.

Alternatively we can choose a different gauge.
For Weyl transformations we still choose $h=0$,
which leaves no ghost term.
For diffeomorphisms we choose the second type of physical
gauge explained in the end of section III.B of \cite{vacca},
namely $\hat{\sigma}=0$ and $\hat{\xi}=0$. 
This is equivalent to taking the Landau gauge limit $a\rightarrow 0$.
The ghosts are a real scalar and a real transverse
vector and the ghost operators are
$\Delta_{L0}-\bR/3$ and $\Delta_{L1}-\bR/2$.
The effective action is given again by (\ref{confdet}).

\subsection{The Einstein-Hilbert case}

Finally we consider the Einstein-Hilbert theory,
where we put $\alpha=\beta=0$.
In this case it is natural to use the two-derivative gauge fixing.
In order to maintain the expressions to a manageable size we write
them here for the case $\omega=m=0$.
Then from the decompositions in sections 2.5 and 2.6
we find that the gauge-invariant variable $h^{TT}$ gives
\be
\Det \Big(\Delta_{L2}-\frac{\br}{2}
+\frac{1}{2}\bar E \Big)^{-1/2}
\ee
the spin-one field $\xi_\mu$ gives
\be
\Det \Big(\Delta_{L1}
+\frac{a-1}{2}\br-2a\Lambda\Big)^{-1/2}
\ee
the scalars $\sigma$ and $h$ give
\be
\Det\left(\Delta_{L0}^2
+\frac{\left(4 \Lambda  \left(-2 a+b^2-3\right)+R \left(2 a-b^2+2 b-3\right)\right)}{(b-3)^2}\Delta_{L0} 
+\frac{4 \Lambda  (4 a \Lambda -a R+R)}{(b-3)^2}\right)
\ee
and finally the ghosts give
\bea
\Det \Big(\Delta_{L1}-\frac{\br}{2} \Big)
\Det \Big(\Delta_{L0}-\frac{\br}{3-b} \Big)\, .
\eea
Putting all together, and making the replacements 
$a\to 1/\gamma$ and $b\to\beta$,
we find that it agrees with equation (4.4) of 
\cite{Fradkin:1983mq}.
If we go on shell by putting $\br=4\Lambda$
the scalar contribution cancels the scalar ghost determinant, leaving just
\bea
\frac
{\sqrt{\Det\left(\Delta_{L1}-\frac{\br}{2}\right)}}
{\sqrt{\Det\left(\Delta_{L2}-\frac{\br}{2}\right)}}\ ,
\eea
which is gauge-independent and agrees with the
classic result of \cite{Christensen:1979iy}.
A slightly more complicated calculation shows that 
also the dependence on the parameters
$\omega$ and $m$ automatically goes away on shell.

\section{The divergences}

\subsection{Derivation}

The one-loop effective action contains a divergent part
\bea
\Gamma_k&=&\int d^dx\,\sqrt{\bg}\bigg[
\frac{A_1}{16\pi d} k^d
+\frac{B_1}{16\pi(d-2)} k^{d-2}\bR
\nonumber\\
&&
+\beta_\alpha\frac{k^{d-4}}{d-4} \bar R^2
+\beta_\beta\frac{k^{d-4}}{d-4} \bR_{\mu\nu}\bR^{\mu\nu}
+\beta_\gamma\frac{k^{d-4}}{d-4} \bR_{\mu\nu\rho\sigma}\bR^{\mu\nu\rho\sigma}
\bigg]\,,
\label{gammaabc}
\eea
where $k$ stands for a cutoff.
In $d=4$, the power divergences $k^{d-4}/(d-4)$
of the last three terms are replaced by terms $\log(k/\mu)$,
where we introduced a reference mass scale $\mu$.
In general one would have separate
Riemann squared, Ricci squared and $R^2$ terms,
but on an Einstein space~(\ref{einstein})
the latter two merge into a single term proportional to $\bR^2$.
Then, we will denote 
\bea
C_1=\beta_\alpha+\frac{1}{d}\beta_\beta ,
\label{quarticcoup}
\eea
the coefficient of the $\bR^2$ divergence and 
$D_1=\beta_\gamma$ the coefficient of the
$\bR_{\mu\nu\rho\sigma}\bR^{\mu\nu\rho\sigma}$
divergence.
We note that in I we used a maximally symmetric background, where
$\bR_{\mu\nu\rho\sigma}\bR^{\mu\nu\rho\sigma}=
\frac{2}{d(d-1)}\bR^2$
Thus, the coefficient $C_1$ of I corresponds to the combination
$C_1+\frac{2}{d(d-1)}D_1$.

We describe here the algorithm that we use to derive the coefficients.
Instead of $\Gamma_k$ we shall evaluate the derivative~\cite{Reuter,Dou}:
\be
\dot\Gamma_k=
\int d^dx\,\sqrt{\bg}\left[
\frac{A_1}{16\pi} k^d
+\frac{B_1}{16\pi} k^{d-2}\bar R
+\left(\beta_\alpha\bar R^2
+\beta_\beta\bR_{\mu\nu}\bR^{\mu\nu}
+\beta_\gamma \bR_{\mu\nu\rho\sigma}\bR^{\mu\nu\rho\sigma}\right)
k^{d-4}
+\ldots\right],
\label{gammadotabc}
\ee
where the overdot stands for $k\frac{d}{dk}$.
The advantage of this procedure is that this quantity is convergent.
Independently of the renormalizability properties of the theory,
it defines a flow on the space of all couplings that are
permitted by the symmetries of the system.
The divergences of (\ref{gammaabc}) in the limit $k\to\infty$
are then found by integrating the differential equation (\ref{gammadotabc}) from some initial scale
$k_0$ up to $k$.
The coefficients $A_1$, $B_1$, $\beta_\alpha$, $\beta_\beta$
and $\beta_\gamma$ enter in the beta functions of the couplings
$\Lambda$, $G$, $\alpha$, $\beta$, $\gamma$,
but we postpone a discussion of this point to section 7.

The technique employed to evaluate the right hand side of 
(\ref{gammadotabc})
is similar to the one described in I, 
but this time the cutoff is taken to be a function of the
Lichnerowicz Laplacians instead of the Bochner Laplacian $-\bnabla^2$,
as in I.
In each spin sector,
the Lichnerowicz operator $\Delta_L$ is replaced by the regularized one
$P_k(\Delta_L)=\Delta_L+R_k(\Delta_L)$,
where we use the optimized cutoff $R_k(\Delta_L)=(k^2-\Delta_L)\t(k^2-\Delta_L)$.
Then $\dot\Gamma_k$ is given by
\be
\dot\Gamma_k=
\frac{1}{2}\Tr\left(\frac{\dot\Delta_{k}^{(2)}}{\Delta_{k}^{(2)}}\right)
+\frac{1}{2}\Tr\left(\frac{\dot\Delta_{k}^{(1)}}{\Delta_{k}^{(1)}}\right)
+\frac{1}{2}\Tr\left(\frac{\dot\Delta_{k}^{(0)}}{\Delta_{k}^{(0)}}\right)
-\Tr\left(\frac{\dot\Delta_{gh,k}^{(1)}}{\Delta_{gh,k}^{(1)}}\right)
-\Tr\left(\frac{\dot\Delta_{gh,k}^{(0)}}{\Delta_{gh,k}^{(0)}}\right)\ ,
\label{gammadotexp}
\ee
where $\Delta_k$ are the kinetic operators 
that appear in the each spin sector
and the numerator is 
$\dot\Delta_{k}=\dot R_k(\lich)=2k^2 \t(k^2-\lich)$.
The step function cuts off the trace to eigenvalues of $\lich$
that are smaller than $k^2$.
For example in the spin-two sector
\bea
\frac{1}{2}\Tr\left(\frac{\dot\Delta_{k}^{(2)}}{\Delta_{k}^{(2)}}\right)
\!\!&=&\!\!
\frac{1}{2}\frac{1}{(4\pi)^{d/2}}\Bigg[
W(\Delta_L^{(2)},0)\left(Q_{d/2}b_0(\Delta_L^{(2)})
+Q_{d/2-1}b_2(\Delta_L^{(2)})
+Q_{d/2-2}b_4(\Delta_L^{(2)})\right)
\nonumber\\
&&
\qquad\qquad
+W'(\Delta_L^{(2)},0)\bR\left(Q_{d/2}b_0(\Delta_L^{(2)})
+Q_{d/2-1}b_2(\Delta_L^{(2)})\right)
\nonumber\\
&&
\qquad\qquad
+\frac{1}{2}W''(\Delta_L^{(2)},0)\bR^2\left(Q_{d/2}b_0(\Delta_L^{(2)})\right)
+\ldots\Bigg],
\label{euterpe}
\eea
where
$W(\Delta_L^{(2)},\bR)=\frac{\dot\Delta_{Lk}^{(2)}}{\Delta_{Lk}^{(2)}}$
and primes denote derivatives with respect to $\bR$.
The coefficients $Q_n$ and the heat kernel coefficients $b_n$ are listed in Appendix A.

Similar formulas hold for the spin one and spin zero sectors and for the ghosts.
In the scalar term, $\Delta_{k}^{(0)}$ is a two-by-two matrix,
and the fraction has to be understood as the product of $\dot\Delta_{k}^{(0)}$
with the inverse of $\Delta_{k}^{(0)}$.
The functional trace thus involves also a trace over this two-by-two matrix.
With these data one can write the expansion of (\ref{gammadotexp})
in powers of $\bR$, and comparing with (\ref{gammadotabc}) one can
read off the coefficients $A_1$, $B_1$ and $C_1$.

\subsection{Results}

From now on we will deal with dimensionless variables 
\begin{equation}
\tilde\alpha = k^{4-d}\alpha\,,\,\,\, \tilde\beta = k^{4-d}\beta\,,\,\,\,
\tilde\Lambda=k^{-2}\Lambda\,,\,\,\,\tilde G = k^{d-2}G\,,\,\,\,
(\tilde Z_N = k^{2-d} Z_N) \,,
\label{dlesscoup}
\end{equation}
and
\bea
\tilde\zgf = \left\{
\begin{tabular}{lll}
$k^{2-d} \zgf$ & for & two-derivative gauge fixing \p{gfghl} \\
$k^{4-d} \zgf$ & for & four-derivative gauge fixing \p{gfgh}
\end{tabular}
\right. .
\eea
In I, the couplings $\tilde\alpha$ and $\tilde\beta$ were absent,
and $\tilde G$ only appeared in the Hessian through the overall prefactor $Z_N$.
Since this prefactor cancelled between numerator and denominator
in (\ref{gammadotexp}), the divergences were independent of $\tilde G$.
This does no longer happen in the present theory, so
the coefficients $A_1$, $B_1$ and $C_1$ depend in general on the dimension $d$,
on the measure parameters $m$ and $\omega$, on the gauge parameters $a$ and $b$,
and on the couplings $\tilde\Lambda$, $\tilde G$, $\tilde\alpha$ and $\tilde\beta$.
The general expressions for the coefficients of the divergences
are extremely complicated and not instructive.
We will therefore only write them in certain limits where they simplify enough.

Compared to I, we have to take into account the additional
dependence on the couplings $\tilde G$, $\tilde\alpha$ and $\tilde\beta$.
We will consider two limiting situations.
One is the limit $\tilde\alpha\to0$ and $\tilde\beta\to0$.
In this case the beta functions should reduce to those of
Einstein-Hilbert gravity and match with those of I
(modulo scheme dependences, due to the different form of
the cutoff in this paper).
Still, the present results are more general because
we consider a generic Einstein background,
which allows us to distinguish at least two of the
higher derivative couplings, whereas in I a maximally
symmetric background was used, allowing us to calculate
the coefficient of a single combination of the higher derivative couplings.
We will refer to this as the ``Einstein-Hilbert limit'' (EH limit).\footnote{
The same results can be obtained, at least in some cases, by taking the limit $\tilde Z_N\to\infty$,
which is the same as considering the perturbative regime $\tilde G\to 0$.}

The other limit consists in taking $\tilde Z_N\to 0$, which is equivalent to $\tilde G\to\infty$.
In this case one is left with functions of the higher derivative couplings only.
We will refer to this as the ``four-derivative gravity limit'' (4DG-limit).

There is still the freedom of choosing between the two-derivative and four-derivative gauges.
In sections 4.3 and 4.4 we will consider the 4DG- and the EH-limits of the theory,
using the two-parameter family of two-derivative gauges introduced in section 2.6.
In section 4.5 we will discuss the changes that occur
when using the two-parameter family of four-derivative gauges introduced in section 2.7.
It will turn out that in order to take the EH- and 4DG-limits, different choices have to be made
regarding the overall gauge-fixing coefficient $Z_{GF}$.
These are spelled out in detail in the following sections.

\subsection{The 4DG-limit $(\tilde Z_N \to 0)$}

For this calculation we set $\tilde\zgf=1$.
The coefficients $A_1$, $B_1$ and $D_1$ turn out to be
universal (i.e. independent of the gauge and parametrization)
in any dimension:
\bea
A_1&=&\frac{8(d-2)}{(4\pi)^{d/2-1}\Gamma(d/2)},
\nonumber
\\
B_1&=&\frac{2}{(4\pi)^{d/2-1}
3d^2\tilde\beta(4(d-1)\tilde\alpha+d\tilde\beta)\Gamma(d/2)}
\big[
48  d \left(d^3-2 d^2-d+2\right)\tilde\alpha^2
\nonumber
\\
&&+4 \left(d^5-12 d^4+41 d^3-102
   d^2+36 d+48\right)\tilde\alpha\tilde\beta
   + \left(d^5-14 d^4+30 d^3-60 d^2-72 d+96\right)\tilde\beta^2
\big],
\nonumber
\\
D_1&=&\frac{-1050 + 589 d - 34 d^2 + d^3}{(4\pi)^{d/2}360\Gamma(d/2)}.
\label{chris}
\eea
The coefficient $C_1$ has a complicated dependence on
$m$, $\omega$, $a$, $b$, $\tilde\alpha$ and $\tilde\beta$, that we do not report here.
A special case will be given below in (\ref{olga}).
However, in $d=4$ all four coefficients, including $C_1$, are universal:
\bea
A_1&=&\frac{4}{\pi},
\nonumber
\\
B_1&=&\frac{15\tilde\alpha-14\tilde\beta}{3\pi\tilde\beta},
\nonumber
\\
C_1&=&\frac{1200\tilde\alpha^2+200\tilde\alpha\tilde\beta
-183\tilde\beta^2}{1920\pi^2 \tilde\beta^2},
\nonumber
\\
D_1&=& \frac{413}{2880\pi^2}.
\label{b1c14d}
\eea
This is an explicit consequence of the statement made in 
section~\ref{quadraticg}, 
that the one-loop effective action of higher derivative gravity
is independent of the parametrization and gauge choice in $d=4$.

The expressions for $C_1$ and $D_1$ given above
are in agreement with the standard results
for the beta functions in HDG \cite{avrabar,BS2}
(also derived by means of (\ref{gammadotexp}) in
\cite{CP,niedermaier,OP}):
\bea
\beta_{\tilde\alpha} &=&
\frac{1}{(4\pi)^2}\frac{90\tilde\alpha^2-23\tilde\beta^2-338\tilde\gamma^2
+15\tilde\alpha\tilde\beta-120\tilde\alpha\tilde\gamma
-199\tilde\beta\tilde\gamma}{9(\tilde\beta+4\tilde\gamma)^2},
\nonumber\\
\beta_{\tilde\beta}&=&\frac{1}{(4\pi)^2}\frac{371}{90},
\nonumber\\
\beta_{\tilde\gamma}&=&\frac{1}{(4\pi)^2}\frac{413}{180}.
\eea

Let us consider what happens if we choose $\tilde\zgf=\tilde Z_N$,
as is usually done in Einstein-Hilbert theory.
In the 4DG-limit, the coefficients $A_1$ and $D_1$ are still given by the formulas given above,
but $B_1$ and $C_1$ are different, and gauge-dependent.
However, for the exponential parametrization ($\omega=\frac12$) in $d=4$, one gets $m$-independent result
\bea
B_1&=&\frac{15\tilde\alpha(b-3)^2+\tilde\beta\left(3a-14(b-3)^2\right)}{3\pi(b-3)^2\tilde\beta},
\label{b1c14d2}
\\
C_1&=&\frac{\tilde\beta ^2 \left(240 a^2+40 a \left(b^2-18 b+45\right)-183 (b-3)^4\right)
+1200 \tilde\alpha^2 (b-3)^4+200 \tilde\alpha \tilde\beta (b-3)^4}{1920 \pi ^2 (b-3)^4 \tilde\beta ^2}.
\nonumber
\eea
We note that in the limit $a\to0$ these reproduce the universal formulas given above.
This observations agrees with the discussion in the end of section 2.6.
In the 4DG-limit, at fixed $k$, one sets $\tilde Z_N=0$.
For fixed $a$ and $k$, and with the choice $\tilde\zgf=\tilde Z_N$,
this implies that the gauge fixing term vanishes too.
The situation can be fixed by letting $a\to0$, with the ratio $\tilde\zgf/a$ fixed and finite.
In conclusion, the gauge-dependence that occurs in (\ref{b1c14d2}) for $a\not=0$ 
is the artifact of a bad gauge-fixing procedure. The correct result is given by (\ref{b1c14d}).

This problem does not arise in the standard calculation
of the beta functions of HDG in $d=4$, because there the gauge fixing is of the type considered
in section \ref{quarticg}. When the gauge-fixing term contains four derivatives,
its overall coefficient is dimensionless in $d=4$ and there
is never the temptation to introduce a factor $\tilde Z_N$.

\subsection{The EH limit}

Let us now consider the EH limit $\tilde\a, \tilde\b \to 0$.
In this case it is not appropriate to set $\tilde\zgf=1$,
as we did  in the preceding subsection,
because of the issue in the gauge fixing
discussed in the end of section 2.6.
Indeed if we insisted with this choice
we would find that for $\tilde\Lambda=0$
only $A_1$ and $D_1$ are universal:
\bea
A_1&=&\frac{4(d-3)}{(4\pi)^{d/2-1}\Gamma(d/2)} ,
\nonumber
\\
D_1&=&\frac{d^3-35d^2+606d-1080}{(4\pi)^{d/2}720\Gamma(d/2)}
\ .
\label{emily1}
\eea
whereas $B_1$ and $C_1$ would contain $\tilde G$ 
and diverge for $\tilde G\to0$.
For this reason we switch now to the choice $\tilde\zgf=\tilde Z_N$,
which is more appropriate to discuss this limit.

We find that $A_1$ and $D_1$ are still given by (\ref{emily1}).
In $d=4$ we have:
\bea
A_1&=&\frac{1}{\pi} ,
\nonumber\\
B_1&=&
\frac{1}{12\pi(b-3)^2(4m+1)}
\Big[
3 a \big(-18 b (4 m+1)^2 (2 \omega -1)
+3 (2 \omega -1) (4 b m+b)^2
\nonumber\\
&&
+8 m (66 (2 m+1) \omega -66
   m-31)
   +70 \omega -31\big)
   +b^2 (4 m+1) (72 m \omega -36 m+18 \omega -47)
\nonumber\\
&&   
   -6 b (4 m+1) (30 (4 m+1)
   \omega -60 m-53)
   +9 (48 m (10 m \omega -5 m+5 \omega -6)+26 \omega -55)
\Big] ,
\nonumber
\\
D_1&=&\frac{53}{720\pi^2} .
\label{ester}
\eea
We note that $\tilde G$ does not appear in these expressions,
so that they have a smooth limit $\tilde G\to 0$.
The expression for $C_1$ is still too long to be written
but is likewise independent of $\tilde G$.
We give it only for two cases: linear parametrization
$m=\omega=0$
\bea
C_1&=&\frac{1}{1920 \pi ^2 (b-3)^4}
\big[
15 a^2 \left(3 b^4-36 b^3+162 b^2-324 b+259\right)
\\
   &&
   -20 a \left(3 b^4-36 b^3+176 b^2-360
   b+297\right)
 +24 \left(7 b^4-59 b^3+223 b^2-381 b+252\right)
\big].
\nonumber
\eea
and exponential parametrization $\omega=1/2$,
in which case the result is automatically independent of $m$:
\be
C_1=\frac{240 a^2+40 a \left(b^2-18 b+9\right)-3 \left(29 b^4-348 b^3+1526 b^2-3372 b+2709\right)}{1920
   \pi ^2 (b-3)^4} .
\ee

One can plot these functions for fixed parametrization 
or for fixed gauge.
One obtains plots that are very similar to those shown in I.
They will not be repeated here.

We report for completeness the expressions $B_1$ and $C_1$ for $d=4$
in the ``unimodular physical'' gauge $b \to\pm\infty$, $a\to0$,
for $\tilde\alpha$, $\tilde\beta$, $\tilde G$ all tending to zero:
\bea
B_1&=&
\frac{36 m (2 \omega -1)+18 \omega -47}{12 \pi },
\nonumber\\
C_1&=&   
\frac{180 m^2 (1-2 \omega )^2+10 m \left(36 \omega ^2-44 \omega +13\right)+45 \omega ^2-65 \omega
   +14}{160 \pi ^2}.
   \label{upg}
\eea

\subsection{Results with the four-derivative gauge fixing}

In the preceding sections we used the two-derivative gauge-fixing
terms introduced in section 2.6. We now discuss briefly the
results when the four-derivative gauge-fixing terms of
section 2.7 are used instead.

The formal discussion of section 3 indicates that in the
4DG-limit in four dimensions, the divergences
should be universal.
Explicit calculation confirms that in this case
the coefficients $A_1$, $B_1$, $C_1$, $D_1$ are indeed given
again by (\ref{b1c14d}).
Furthermore, the coefficients $A_1$, $B_1$ and $D_1$
are universal in any dimension and agree with those
given in (\ref{chris}).

In the EH limit the coefficients $A_1$, and $D_1$
in $d=4$ are again as in (\ref{ester}), but the others are 
slightly different.
For $B_1$ one has in general
\bea
B_1&=&
\frac{1}{12\pi (b-3)^2 (4 m+1)}
\Big[
6 a \big(-18 b (4 m+1)^2 (2 \omega -1)+3 (2 \omega -1) (4 b m+b)^2
\nonumber\\
&&+8 m (66 (2 m+1) \omega -66 m-31)
+70\omega-31\big)
+b^2 (4 m+1) (72 m \omega -36 m+18 \omega -47)
\nonumber\\
&&-6 b (4 m+1)(30(4m+1)\omega-60m-53)
+9 (48 m (10 m \omega -5 m+5 \omega -6)+26 \omega -55)
\Big]\, . \nonumber\\,
\eea
Similarly the coefficient $C_1$ in the linear parametrization 
$m=\omega=0$ is
\bea
C_1&=&\frac{1}{1920 \pi ^2 (b-3)^4}
\big[
30 a^2 \left(3 b^4-36 b^3+162 b^2-324 b+259\right)
\\
&&
-5a\left(15b^4-180b^3+898b^2-1860b+1575\right)
+24 \left(7 b^4-59 b^3+223 b^2-381 b+252\right)
\big].
\nonumber
\eea
and in the exponential parametrization $\omega=1/2$:
\be
C_1=\frac{480a^2+80a\left(b^2-15b+9\right)
-3\left(29b^4-348b^3+1526b^2-3372 b+2709\right)}{1920\pi^2(b-3)^4} .
\ee
The result in the unimodular physical gauge $b \to\pm\infty$, $a\to0$
agrees with (\ref{upg}).

\subsection{Exponential parametrization and unimodular gauge}

It was found in I that choosing the exponential parametrization
$\omega=1/2$ and the unimodular gauge $b\to\pm\infty$,
all dependence on the other two parameters $m$ and $a$,
as well as the dependence on the cosmological constant $\tilde\Lambda$,
drop out.

This result holds true also in the present context,
in any dimension and independently of whether the gauge fixing
contains two or four derivatives and independently of
the treatment of the constant $\tilde\zgf$.
The resulting coefficients are still too cumbersome to write,
so we give them again only in two limits.

In the 4DG-limit the coefficients $A_1$, $B_1$ and $D_1$
are given by (\ref{chris}) and
\bea
C_1&=&\frac{1}{720(4\pi)^{d/2}d^3\tilde\beta^2
(4(d-1)\tilde\alpha+d\tilde\beta)^2\Gamma(d/2)}
\Big[
46080\,\tilde\alpha^4 (d-1)^2 d^2 \left(d^2-d-2\right)
\\
&&+1920\,\tilde\alpha^3\,\tilde\beta d
\left(d^6-3 d^5+85 d^4-285 d^3+106 d^2+288 d-192\right)
\nonumber\\
&&+16\,\tilde\alpha^2\tilde\beta^2\left(5d^8-92d^7+917d^6-1886d^5+7964d^4-49268 d^3+42360d^2+25920d-23040\right)
\nonumber\\
&&+8\,\tilde\alpha\tilde\beta^3d\left(5d^7-132d^6+1235d^5-4716d^4+10988d^3-28800 d^2-14880d+46080\right)
\nonumber\\
&&+\tilde\beta^4 \left(5 d^8-142 d^7+1288 d^6-4628 d^5+7560 d^4-19200 d^3-17280
d^2-46080 d+92160\right)
\Big]\ .
\nonumber
\label{olga}
\eea

In the EH limit we have
\bea
A_1&=&\frac{4(d-3)}{(4\pi)^{d/2-1}\Gamma(d/2)} ,
\nonumber
\\
B_1&=&\frac{(d^3-15d^2+24d-72)}{3d(4\pi)^{d/2-1}\Gamma(d/2)} ,
\nonumber
\\
C_1&=&\frac{5d^5-147d^4+1240d^3-3612d^2+2880d-17280}{1440d^2(4\pi)^{d/2}\Gamma(d/2)}
\nonumber\\
D_1&=&\frac{d^3-35d^2+606d-1080}{720(4\pi)^{d/2}\Gamma(d/2)}
\ .
\label{emily}
\eea
We note that the formulas for $B_1$ and $C_1+\frac{2}{d(d-1)}D_1$
do not agree with equation (5.7) in I.
This is due to the different choice of cutoff (a function of the
Bochner Laplacian in I and of the Lichnerowicz Laplacians here).
Specializing to $d=4$, the formula for $C_1+\frac{2}{d(d-1)}D_1$ 
agrees with (5.6) in I but $B_1$ does not. This is a signal of
the lack of universality of this coefficient.

\section{Conformal gravity in $d=4$}

Independently of the choice of gauge and parametrization,
the effective action for conformal gravity is
given formally by (\ref{confdet}).
This leads to the following flow for the effective action
\be
\dot\Gamma_k=
\frac{1}{2}\Tr\left(\frac{\dot\Delta_{k}^{(2)}}{\Delta_{k}^{(2)}}\right)
-\frac{1}{2}\Tr\left(\frac{\dot\Delta_{gh,k}^{(1)}}{\Delta_{gh,k}^{(1)}}\right)
-\frac{1}{2}\Tr\left(\frac{\dot\Delta_{gh,k}^{(0)}}{\Delta_{gh,k}^{(0)}}\right)
\ ,
\label{gammadotconf}
\ee
where 
\bea
\Delta_k^{(2)}&=&
\left(P_k(\lich_2)-\frac{1}{2}\bR\right)
\left(P_k(\lich_2)-\frac{1}{3}\bR\right)\ ,
\nonumber\\
\Delta_k^{(1)}&=&P_k(\lich_1)-\frac{1}{2}\bR\ ,
\nonumber\\
\Delta_k^{(0)}&=&P_k(\lich_0)-\frac{1}{3}\bR\ .
\eea
Proceeding as before we find
\bea
A_1&=&\frac{3}{(4\pi)^2}\ ,
\\
B_1&=&\frac{41}{6(4\pi)^2}\ ,
\\
C_1&=&-\frac{199}{180(4\pi)^2}\ ,
\\
D_1&=&\frac{137}{60(4\pi)^2}\ .
\eea
We observe that the logarithmic divergences are given by
\bea
\frac{1}{(4\pi)^2}\left[
b_4\left(\Delta_{L2}-\frac{1}{2}\bR\right)
+b_4\left(\Delta_{L2}-\frac{1}{3}\bR\right)
-b_4\left(\Delta_{L1}-\frac{1}{2}\bR\right)
-b_4\left(\Delta_{L2}-\frac{1}{2}\bR\right)\right]\, .
\label{conflogdiv}
\eea
The coefficients $C_1$ and $D_1$ agree with 
previous calculations reported in \cite{BS1} and \cite{pang}.

\section{Duality}

One of the main results of I was the existence of a discrete
idempotent transformation leaving the functional measure,
and the divergences, invariant.
That invariance extends also to HDG, at least on an Einstein background.
In any dimension and in any gauge we find that the
divergences $A_1, B_1$, $C_1$ and $D_1$ have the duality symmetry:
\bea
A_1(\omega,m)&=&A_1\left(1-\omega,-m-\frac{2}{d}\right)\ ,
\nonumber\\
B_1(\omega,m)&=&B_1\left(1-\omega,-m-\frac{2}{d}\right)\ ,
\nonumber\\
C_1(\omega,m)&=&B_1\left(1-\omega,-m-\frac{2}{d}\right)\ ,
\nonumber\\
D_1(\omega,m)&=&C_1\left(1-\omega,-m-\frac{2}{d}\right)\ .
\label{duality}
\eea

This duality is only manifest if the coefficient $b$
enters in the gauge fixing (\ref{gaugecondition})
through the combination $\bar b=b(1+dm)$.
With other definitions of this coefficient a form of duality will
still be present but it will have a much more complicated form.
The calculations reported here indicate that this duality
is not limited to Einstein gravity.
It will be interesting to understand more generally
under what circumstances this property holds.

A transformation can be an invariance of a quantum theory
if it leaves invariant the action (and its expansion)
and the functional measure. 
As discussed in the introduction, in our one-loop calculations
we keep the functional measure fixed to be (\ref{measure})
and the origin of the
$\omega$- and $m$-dependence of the results must lie in the
form of the Hessians. 
One can indeed check explicitly that the
Hessians given in sections 2.5-2.6 are duality-invariant.

Furthermore, it was observed in I that the functional measures 
$$
\Pi_x d\gamma_{\mu\nu}(x)=\Pi_x d\hat h_{\mu\nu}(x)
$$
where $\gamma_{\mu\nu}$ and $\hat h_{\mu\nu}$ have weight $w$, and
$$
\Pi_x d\gamma^{\mu\nu}(x)=\Pi_x d\hat h^{\mu\nu}(x)
$$
where $\gamma^{\mu\nu}$ and $\hat h^{\mu\nu}$ have weight $w'$, 
related by
\be
\frac{w'}{2}=\frac{w}{2}+\frac{2}{d}\ ,
\quad\mathrm{which\ is\ the\ same\ as}\quad
m'=-m-\frac{2}{d}\ ,
\ee
are equivalent.
Thus, if we keep the same action, the functional integrals
are equivalent.
Again, these measures would give rise to different power-law divergences coefficients, but duality would still hold for each choice of measure.

Conversely, we could understand the invariance of the hessian as follows:
We have the functional integral
\bea
\int [\Pi_x d\gamma_{\mu\nu}(x)] e^{-S[g_{\mu\nu}]},
\label{fi1}
\eea
where $\c_{\mu\nu}$ is related to the metric by~\cite{I}
\bea
\c_{\mu\nu} = g_{\mu\nu}(\det g)^{w/2}.
\label{gamma}
\eea
Solving \p{gamma} for $g_{\mu\nu}$ and substituting it into \p{fi1}, we have 
\bea
\int [\Pi_x d\gamma_{\mu\nu}(x)] e^{-S[\c_{\mu\nu}(\det \c)^{m}]},
\eea
with $m=-\frac{w/2}{1+dw/2}$.
Simply rewriting $\c_{\mu\nu}$ as $g_{\mu\nu}$, we are lead to
\bea
\int [\Pi_x dg_{\mu\nu}(x)] e^{-S[g_{\mu\nu}(\det g)^{m}]}.
\eea
This is a functional integral with fixed quantum field, but the metric in the action is transformed.
Viewed this way, the invariance should appear as an invariance of the action with the quantum field kept fixed.
This is precisely the calculation we have given. The above discussion by the invariance of the measure
suggests that duality is exact, but our calculation is done only at one loop.
It would be interesting to check if the latter approach also gives exact result.

The existence of the duality is, however, more general.
If we used a functional measure that contains the determinant
of a ``De Witt'' metric in functional space,
the measure itself would be invariant under arbitrary
field redefinitions \cite{Mottola:1995sj}.
In particular, it would be independent of $m$ and of the
choices $\omega=0,1/2,1$.
Such a measure would give rise to different power-law divergences coefficients 
than the ones reported here, but duality would again appear because 
it is an invariance of the Hessian.

\section{Concluding remarks}

In this paper we have extended the analysis of I \cite{I}
from Einstein-Hilbert gravity to higher-derivative gravity
containing the squares of the Ricci scalar and Ricci tensor.
In four-dimensions the analysis is essentially complete,
because the remaining independent invariant is a total derivative.
In higher dimensions this is not so.
The analysis was also limited to Einstein backgrounds,
which nearly solve the equations of motion,
but also in this way is more general than the analysis in I,
which was limited to maximally symmetric backgrounds.

We have obtained formulas for the one-loop divergences up to
quadratic terms in the curvature.
These are all the divergences that arise at one loop in $d=4$.
The method we have used is a one-loop approximation
of the single-field approximation of the Exact RG equation
for gravity, as first derived in \cite{Reuter} for the
EH-case and then extended to HDG in \cite{CP,niedermaier,OP},
to 3-$d$ topologically massive gravity in \cite{ps},
and beyond the one-loop approximation in
\cite{lauscher,bms,sgrz}.

The coefficients are related to the beta functions for the couplings.
Comparing the action~\p{action} with our results~\p{gammadotabc}, we find the beta functions
for the dimensionless couplings~\p{dlesscoup} as
\bea
\b_{\tilde G} &=& (d-2)\tilde G + B_1 \tilde G^2, \nn
\b_{\tilde \Lambda} &=& -2 \tilde\Lambda + B_1 \tilde\Lambda \tilde G +\frac{A_1}{2} \tilde G, \nn
\b_{\tilde\a+\frac{1}{d}\tilde\b} &=& -(d-4) \left(\tilde\a+\frac{1}{d}\tilde\b\right) + C_1.
\eea
As already remarked, working on an Einstein space
prevents us from disentangling the beta functions of $\tilde\alpha$
and $\tilde\beta$.
On the other hand, even though we did not write a term
$\gamma R_{\mu\nu\rho\sigma}R^{\mu\nu\rho\sigma}$
in the action, the divergence $D_1$ gives rise to a beta function
\be
\b_{\tilde\gamma} = -(d-4)\tilde\gamma + D_1.
\ee
The signs of the coefficients imply that $\alpha$, $\beta$, $\gamma$
are asymptotically free in $d=4$  (for suitable range of parameters)~\cite{julve,ft1,avrabar}
whereas $\tilde G$  goes in the UV to a fixed point $\tilde G = -(d-2)/B_1$.
In order for this nontrivial fixed point to make sense, $B_1$ must be negative.
In spite of the non-universality of the beta functions,
many calculations indicate that $B_1$ is indeed
negative for the linear split.
For example, $B_1$ in \p{ester} is negative for $m=\omega=0$ and a wide range of $a$ and $b$.
With the exponential parametrization discussed in subsection~4.6, $B_1$ in the EH limit
given in \p{emily} is negative for $3 \leq d \leq 13$.
We refer the reader to the literature for a more detailed discussion.

The explicit coefficients of the divergences, 
in arbitrary dimension, gauge and parametrization, 
are too complicated to write,
and we have exhibited only some special cases.
The universal values agree with the literature.
The divergences also agree, in $d=4$ and in the EH limit,
with earlier calculations in general gauges \cite{kallosh,Fradkin:1983mq,Kalmykov:1995fd,Gies:2015tca}.
For further discussions on the use of the
exponential parametrization see \cite{vacca,nink}.

Most of the general features noted in I persist in the theories
considered here, as we have observed.
In particular,
one striking feature that we have checked in full generality is the
existence of a ``duality'' symmetry under the change 
of parametrization (\ref{dudu}), or
\bea
\omega\to 1-\omega\ ;\qquad m\to -m-\frac{2}{d}\ .
\eea
As observed in I, this transformation preserves 
the dimension of the quantum field
and is also an invariance of the functional measure.
\footnote{We recall that the pairs of measures
proposed in \cite{FY} are dual in this sense.}
However, the parametrization-dependence of the divergences is not due to the actual choice of the
functional measure.
Instead, it comes entirely from the different form of the Hessians
in different parametrizations.
Since these Hessians only differ by terms that are
proportional to the equations of motion,
the parametrization-dependence, as well as the gauge-dependence,
goes away on shell.

Different choices of ultralocal functional measure would
alter the results for the power-law divergences coefficients.
We refer to \cite{Toms:1986sh} and references quoted therein
for a discussion of this point
and to \cite{anselmi} for more general results
using Pauli-Villars regularization.

Whether the duality extends also to other classes of actions,
to other backgrounds and to higher loops are all questions
that we leave for further investigation.
Also left for future work is the calculation of divergences
in the unimodular case $m=-1/d$,
which contains in particular the unique self-dual theory
$\omega=1/2$, $m=-1/d$.

\section*{Acknowledgment}

This work was supported in part by the Grant-in-Aid for
Scientific Research Fund of the JSPS (C) No. 16K05331. ADP is grateful to CNPq for
financial support.

\appendix

\section{$Q_n$ and Heat kernel coefficients for Lichnerowicz Laplacians}
\label{heat}

Here we list the coefficients used in subsection~4.1.
The coefficients $Q_n$ are defined by
\bea
Q_n = \frac{k^{2n}}{\G(n)} \int_0^1 y^{n-1} dy = \frac{k^{2n}}{\G(n+1)}.
\eea

The heat kernel coefficients for the Lichnerowicz Laplacians
acting on spin-0, spin-1 and spin-2 fields 
on an Einstein manifold are,
\bea
b_0(\Delta_L^{(0)}) &=& 1,
\nn
b_2(\Delta_L^{(0)}) &=& \frac{1}{6}R,
\nn
b_4(\Delta_L^{(0)}) &=& \frac{1}{180}R_{\mu\nu\rho\sigma}R^{\mu\nu\rho\sigma}
+\frac{5d-2}{360d}R^2 ,
\\
b_0(\Delta_L^{(1)}) &=& d-1,
\nn
b_2(\Delta_L^{(1)}) &=& \frac{d-7}{6}R,
\nn
b_4(\Delta_L^{(1)}) &=& \frac{d-16}{180} R_{\mu\nu\rho\sigma}R^{\mu\nu\rho\sigma}
+\frac{5d^2-67d+182}{360d}R^2,
\\
b_0(\Delta_L^{(2)}) &=& \frac{(d+1)(d-2)}{2},
\nn
b_2(\Delta_L^{(2)}) &=& \frac{d^2-13d-14}{12}R,
\nn
b_4(\Delta_L^{(2)}) &=& \frac{d^2-31d+508}{360} R_{\mu\nu\rho\sigma}R^{\mu\nu\rho\sigma}
+\frac{5d^3-127d^2+592d+1804}{720d}R^2.
\eea
These formulae can be obtained by the methods described
for example in Appendix B of \cite{CPR}.
Here we do {\it not} take into account isolated
modes that do not contribute to the fluctuation $h_{\mu\nu}$.
Thus these formulae hold in the case when the manifold has no 
Killing or conformal Killing vectors, 
or else if the manifold is noncompact and has
a continuous spectrum, so that the spurious isolated modes
are of measure zero.

In order to get the heat kernel coefficients
of shifted Lichnerowicz Laplacians $\Delta+aR$, one can use
\bea
b_0(\Delta+aR)&=&b_0(\Delta),
\nn
b_2(\Delta+aR)&=&b_2(\Delta)-aRb_0(\Delta),
\nn
b_4(\Delta+aR)&=&b_4(\Delta)-aRb_2(\Delta)+\frac{1}{2}a^2R^2 b_0(\Delta).
\label{heatrel}
\eea


\begin{thebibliography}{99}

\bibitem{I}
N.~Ohta, R.~Percacci and A.~D.~Pereira,
``Gauges and functional measures in quantum gravity I: Einstein theory,''
JHEP {\bf 1606} (2016) 115
[arXiv:1605.00454 [hep-th]].

\bibitem{Modesto:2011kw}
L.~Modesto,
``Super-renormalizable Quantum Gravity,''
Phys.\ Rev.\ D {\bf 86} (2012) 044005
[arXiv:1107.2403 [hep-th]].

\bibitem{Biswas:2011ar}
T.~Biswas, E.~Gerwick, T.~Koivisto and A.~Mazumdar,
``Towards singularity and ghost free theories of gravity,''
Phys.\ Rev.\ Lett.\  {\bf 108} (2012) 031101
[arXiv:1110.5249 [gr-qc]].

\bibitem{Modesto:2014lga}
L.~Modesto and L.~Rachwal,
``Super-renormalizable and finite gravitational theories,''
Nucl.\ Phys.\ B {\bf 889} (2014) 228
[arXiv:1407.8036 [hep-th]].

\bibitem{Stelle1}
K.~S.~Stelle,
``Renormalization of Higher Derivative Quantum Gravity,''
Phys.\ Rev.\  D {\bf 16} (1977) 953.

\bibitem{julve}
J. Julve, M. Tonin,
 ``Quantum Gravity with Higher Derivative Terms,''
  Nuovo Cim.\ B {\bf 46} (1978) 137.

\bibitem{ft1}
E.S. Fradkin, A.A. Tseytlin,
``Renormalizable Asymptotically Free Quantum Theory Of Gravity,''
  Phys.\ Lett.\ B {\bf 104} (1981) 377;
``Renormalizable asymptotically free quantum theory of gravity,''
 Nucl.\ Phys.\ B {\bf 201} (1982) 469.

\bibitem{avrabar}
I.~G.~Avramidi and A.~O.~Barvinsky,
 ``Asymptotic Freedom In Higher Derivative Quantum Gravity,''
Phys.\ Lett.\ B {\bf 159} (1985) 269.\\
I.~G.~Avramidi,
``Covariant methods for the calculation of the effective action in quantum field theory
and investigation of higher derivative quantum gravity,''
hep-th/9510140.
  
\bibitem{salam} 
A.~Salam and J.~A.~Strathdee,
``Remarks On High-Energy Stability And Renormalizability Of Gravity Theory,''
Phys.\ Rev.\  D {\bf 18} (1978) 4480.

\bibitem{floper4}
R.~Floreanini and R.~Percacci,
``The Renormalization group flow of the Dilaton potential,''
Phys.\ Rev.\ D {\bf 52} (1995) 896
[hep-th/9412181].
  
\bibitem{bonannoreuter1}
A.~Bonanno and M.~Reuter,
``Modulated Ground State of Gravity Theories with Stabilized Conformal Factor,''
Phys.\ Rev.\ D {\bf 87} (2013) 084019
[arXiv:1302.2928 [hep-th]].

\bibitem{Shapiro:2015uxa} 
I.~L.~Shapiro,
``Counting ghosts in the “ghost-free” non-local gravity,''
Phys.\ Lett.\ B {\bf 744}, 67 (2015)
[arXiv:1502.00106 [hep-th]].
  
\bibitem{Tomboulis}
E.~Tomboulis,
``1/N Expansion And Renormalization In Quantum Gravity,''
Phys.\ Lett.\  B {\bf 70} (1977) 361.
\\
E.~Tomboulis,
``Renormalizability And Asymptotic Freedom In Quantum Gravity,''
Phys.\ Lett.\  B {\bf 97} (1980) 77.
\\
E.~T.~Tomboulis,
``Unitarity in Higher Derivative Quantum Gravity,''
Phys.\ Rev.\ Lett.\  {\bf 52} (1984) 1173.

\bibitem{Mannheim:2006rd}
P.~D.~Mannheim,
``Solution to the ghost problem in fourth order derivative theories,''
Found.\ Phys.\  {\bf 37} (2007) 532
[hep-th/0608154].

\bibitem{Mukohyama:2013ew}
S.~Mukohyama and J.~P.~Uzan,
``From configuration to dynamics: Emergence of Lorentz signature in classical field theory,''
Phys.\ Rev.\ D {\bf 87} (2013) 065020
[arXiv:1301.1361 [hep-th]].

\bibitem{Salvio:2014soa}
A.~Salvio and A.~Strumia,
``Agravity,''
JHEP {\bf 1406} (2014) 080
[arXiv:1403.4226 [hep-ph]].

\bibitem{Einhorn:2014gfa}
M.~B.~Einhorn and D.~R.~T.~Jones,
``Naturalness and Dimensional Transmutation in Classically Scale-Invariant Gravity,''
JHEP {\bf 1503} (2015) 047,
[arXiv:1410.8513 [hep-th]].
  
\bibitem{AKKLR}
L.~Alvarez-Gaume, A.~Kehagias, C.~Kounnas, D.~L\"ust and A.~Riotto,
``Aspects of Quadratic Gravity,''
Fortsch.\ Phys.\  {\bf 64} (2016) 176
[arXiv:1505.07657 [hep-th]].

\bibitem{KP}
K.~A.~Kazakov and P.~I.~Pronin,
  ``Gauge and parametrization dependence in higher derivative quantum gravity,''
  Phys.\ Rev.\ D {\bf 59} (1999) 064012
  [hep-th/9806023].

\bibitem{OP}
N.~Ohta and R.~Percacci,
``Higher Derivative Gravity and Asymptotic Safety in Diverse Dimensions,''
Class.\ Quant.\ Grav.\  {\bf 31} (2014) 015024
[arXiv:1308.3398 [hep-th]].

\bibitem{HKO}
H.~Hata, T.~Kugo and N.~Ohta,
``Skew Symmetric Tensor Gauge Field Theory Dynamically Realized in {QCD} U(1) Channel,''
Nucl.\ Phys.\ B {\bf 178} (1981) 527.\\
T.~Kugo and S.~Uehara,
``General Procedure of Gauge Fixing Based on {BRS} Invariance Principle,''
Nucl.\ Phys.\ B {\bf 197} (1982) 378.

\bibitem{vacca}
R.~Percacci and G.~P.~Vacca,
``Search of scaling solutions in scalar-tensor gravity,''
Phys.Rev. D92, 061501 (2015)
arXiv:1501.00888 [hep-th].

\bibitem{ppps}
R.~Percacci, M.~J.~Perry, C.~N.~Pope and E.~Sezgin,
``Beta Functions of Topologically Massive Supergravity,''
JHEP {\bf 1403} (2014) 083
[arXiv:1302.0868 [hep-th]].

\bibitem{Fradkin:1983mq}
E.~S.~Fradkin and A.~A.~Tseytlin,
``One Loop Effective Potential in Gauged O(4) Supergravity,''
Nucl.\ Phys.\ B {\bf 234}, 472 (1984).

\bibitem{Christensen:1979iy}
S.~M.~Christensen and M.~J.~Duff,
``Quantizing Gravity with a Cosmological Constant,''
Nucl.\ Phys.\ B {\bf 170} (1980) 480.

\bibitem{Reuter}
M.~Reuter,
``Nonperturbative evolution equation for quantum gravity,''
Phys.\ Rev.\ D {\bf 57} (1998) 971
[hep-th/9605030].

\bibitem{Dou}
D.~Dou and R.~Percacci,
``The running gravitational couplings,''
Class.\ Quant.\ Grav.\  {\bf 15} (1998) 3449
[hep-th/9707239].

\bibitem{BS2}
G.~de Berredo-Peixoto and I.~L.~Shapiro,
``Higher derivative quantum gravity with Gauss-Bonnet term,''
Phys.\ Rev.\ D {\bf 71} (2005) 064005
[hep-th/0412249].
 
\bibitem{CP}
A.~Codello and R.~Percacci,
``Fixed points of higher derivative gravity,''
Phys.\ Rev.\ Lett.\  {\bf 97} (2006) 221301
[hep-th/0607128].
 
\bibitem{niedermaier}
M.~Niedermaier,
``Gravitational Fixed Points from Perturbation Theory,''
Phys.\ Rev.\ Lett.\  {\bf 103} (2009) 101303;
``Gravitational fixed points and asymptotic safety from perturbation theory,''
Nucl.\ Phys.\ B {\bf 833} (2010) 226.

\bibitem{BS1}
G.~de Berredo-Peixoto and I.~L.~Shapiro,
``Conformal quantum gravity with the Gauss-Bonnet term,''
Phys.\ Rev.\ D {\bf 70} (2004) 044024
[hep-th/0307030].

\bibitem{pang}
Y.~Pang,
``One-Loop Divergences in 6D Conformal Gravity,''
Phys.\ Rev.\ D {\bf 86} (2012) 084039
[arXiv:1208.0877 [hep-th]].

\bibitem{Mottola:1995sj} 
E.~Mottola,
``Functional integration over geometries,''
J.\ Math.\ Phys.\  {\bf 36} (1995) 2470
[arXiv:hep-th/9502109].

\bibitem{ps}
R.~Percacci and E.~Sezgin,
``One Loop Beta Functions in Topologically Massive Gravity,''
Class.\ Quant.\ Grav.\  {\bf 27 } (2010)  155009
[arXiv:1002.2640 [hep-th]]

\bibitem{lauscher}
O.~Lauscher and M.~Reuter,
``Flow equation of quantum Einstein gravity in a higher derivative truncation,''
Phys.\ Rev.\ D {\bf 66} (2002) 025026
[hep-th/0205062].

\bibitem{bms}
D.~Benedetti, P.~F.~Machado and F.~Saueressig,
``Asymptotic safety in higher-derivative gravity,''
Mod.\ Phys.\ Lett.\ A {\bf 24} (2009) 2233
[arXiv:0901.2984 [hep-th]];
``Taming perturbative divergences in asymptotically safe gravity,''
Nucl.\ Phys.\ B {\bf 824} (2010) 168
[arXiv:0902.4630 [hep-th]].
  
\bibitem{sgrz}
K.~Groh, S.~Rechenberger, F.~Saueressig and O.~Zanusso,
``Higher Derivative Gravity from the Universal Renormalization Group Machine,''
PoS EPS {\bf -HEP2011} (2011) 124
[arXiv:1111.1743 [hep-th]].

\bibitem{kallosh}
R.~E.~Kallosh, O.~V.~Tarasov and I.~V.~Tyutin,
``One Loop Finiteness Of Quantum Gravity Off Mass Shell,''
Nucl.\ Phys.\ B {\bf 137} (1978) 145.

\bibitem{Kalmykov:1995fd}
M.~Y.~Kalmykov,
``Gauge and parametrization dependencies of the one loop counterterms in the Einstein gravity,''
Class.\ Quant.\ Grav.\  {\bf 12} (1995) 1401
[hep-th/9502152].\\
%
M.~Y.~Kalmykov, K.~A.~Kazakov, P.~I.~Pronin and K.~V.~Stepanyantz,
``Detailed analysis of the dependence of the one loop counterterms on the gauge and parametrization in the Einstein
gravity with the cosmological constant,''
Class.\ Quant.\ Grav.\  {\bf 15} (1998) 3777
[hep-th/9809169].

\bibitem{Gies:2015tca}
H.~Gies, B.~Knorr, S.~Lippoldt,
``Generalized Parametrization Dependence in Quantum Gravity,''
Phys.\ Rev.\ D {\bf 92} (2015) 084020, 
[arXiv:1507.08859 [hep-th]].

\bibitem{nink}
A.~Nink,
``Field Parametrization Dependence in Asymptotically Safe Quantum Gravity,''
Phys.\ Rev.\ D {\bf 91} (2015) 044030
[arXiv:1410.7816 [hep-th]]. \\
M.~Demmel and A.~Nink,
Phys.\ Rev.\ D {\bf 92} (2015) 104013
[arXiv:1506.03809 [gr-qc]].\\
N.~Ohta and R.~Percacci,
``Ultraviolet Fixed Points in Conformal Gravity and General Quadratic Theories,''
Class.\ Quant.\ Grav.\  {\bf 33} (2016) 035001
[arXiv:1506.05526 [hep-th]].
\\ 
N.~Ohta, R.~Percacci and G.~P.~Vacca,
``Flow equation for $f(R)$ gravity and some of its exact solutions,''
Phys.\ Rev.\ D {\bf 92} (2015) 061501
[arXiv:1507.00968 [hep-th]];
\\
``Renormalization Group Equation and scaling solutions for f(R) gravity in exponential parametrization,''
Eur.\ Phys.\ J.\ C {\bf 76} (2016) 46
[arXiv:1511.09393 [hep-th]].
\\
K.~Falls and N.~Ohta,
  ``Renormalization Group Equation for $f(R)$ gravity on hyperbolic spaces,''
  Phys.\ Rev.\ D {\bf 94} (2016) 084005
  [arXiv:1607.08460 [hep-th]].

\bibitem{FY}
K.~Fujikawa,
``Path Integral Measure for Gravitational Interactions,''
Nucl.\ Phys.\ B {\bf 226} (1983) 437.\\
K.~Fujikawa and O.~Yasuda,
``Path Integral for Gravity and Supergravity,''
Nucl.\ Phys.\ B {\bf 245} (1984) 436.

\bibitem{Toms:1986sh}
D.~J.~Toms,
``The Functional Measure for Quantum Field Theory in Curved Space-time,''
Phys.\ Rev.\ D {\bf 35} (1987) 3796.

\bibitem{anselmi}
D.~Anselmi,
``Functional integration measure in quantum gravity,''
Phys.\ Rev.\ D {\bf 45} (1992) 4473.
``On delta(0) divergences and the functional integration measure,''
Phys.\ Rev.\ D {\bf 48} (1993) 680.

\bibitem{CPR}
A.~Codello, R.~Percacci and C.~Rahmede,
``Investigating the Ultraviolet Properties of Gravity with a Wilsonian
Renormalization Group Equation,''
Annals Phys.\  {\bf 324} (2009) 414
[arXiv:0805.2909 [hep-th]].





%
%
%
%
%
%
%
%
%
%


\end{thebibliography}
\end{document}